\documentclass[final]{siamltex}

\usepackage{amsmath}
\usepackage{amsfonts}
\usepackage{mathrsfs}
\usepackage{graphicx}
\usepackage{amssymb}
\usepackage{latexsym}
\usepackage{array}
\usepackage{ifthen}
\usepackage{wrapfig}
\usepackage{picinpar}
\usepackage{rotating}
\usepackage{morefloats}
\usepackage{enumerate}
\usepackage{bm}
\usepackage{stfloats}
\usepackage{dsfont}
\usepackage[english]{babel}
\usepackage{color}
\usepackage{subfigure}
\DeclareMathOperator{\sgn}{sgn}

\newtheorem{example}{Example}

\title{A Phase Theory of MIMO LTI Systems
\thanks{This work was supported in part by National Natural Science Foundation of China under grants 62073003, 72131001, and Hong Hong Research Grants Council under project number GRF 16200619. Some preliminary results were presented in the 58th IEEE Conference on Decision and Control, Nice, France, 2019.}}

\author{Wei Chen\thanks{Department of Mechanics and Engineering Science \& State Key Laboratory for Turbulence and Complex Systems, Peking University, Beijing, China (email: w.chen@pku.edu.cn).}
\and Dan Wang\thanks{Department of Electronic and Computer Engineering, The Hong Kong University of Science and Technology, Clear Water Bay, Kowloon, Hong Kong, China (email: dwangah@connect.ust.hk, eeqiu@ust.hk).} \and Sei Zhen Khong\thanks{Independent scholar (email: szkhongwork@gmail.com).}
\and Li Qiu\footnotemark[3]}

\begin{document}

\maketitle

\begin{abstract}
In this paper, we define the phase response for a class of multi-input multi-output (MIMO) linear time-invariant (LTI) systems whose frequency responses are (semi-)sectorial at all frequencies. The newly defined phase concept subsumes the well-known notions of positive real systems and negative imaginary systems. We formulate a small phase theorem for feedback stability, which complements the celebrated small gain theorem. The small phase theorem lays the foundation of a phase theory of MIMO systems. We also discuss time-domain interpretations of phase-bounded systems via both energy signal analysis and power signal analysis. In addition, a sectored real lemma is derived for the computation of MIMO phases, which serves as a natural counterpart of the bounded real lemma.
\end{abstract}



\pagestyle{myheadings}
\thispagestyle{plain}
\markboth{WEI CHEN, DAN WANG, SEI ZHEN KHONG and LI QIU}{A PHASE THEORY OF MIMO LTI SYSTEMS}

\section{Introduction}
In the classical frequency domain analysis of single-input-single-output (SISO) LTI systems, the magnitude (gain) response and phase response go hand in hand. In particular, the
Bode magnitude and phase plots are always drawn shoulder to shoulder. The combined Bode plot of a loop transfer function provides a significant amount of useful information about the closed-loop stability and performance. The gain and phase crossover frequencies of a loop transfer function give salient information on the
gain and phase margins of the feedback system, which are two useful tools widely used for characterizing the closed-loop
stability and performance. The famous Bode gain-phase integral relation binds the gain and phase together. In frequency domain controller synthesis, phase also plays an important role. Loop-shaping design techniques, such as lead and lag compensation, are rooted in the phase stabilization ideas. Therefore, the gain and phase concepts constitute
the two equally important interrelated pillars of SISO LTI system control theory.

The inception of MIMO system theory sees extension and thriving of the magnitude concept, but not equal flourishing in the phase concept. How to define the phases of MIMO systems has remained mysterious for more than half a century. An early attempt was made in \cite{Macfarlane1981} by considering phases of eigenvalues of the unitary part of the matrix polar decomposition. Later, the authors in \cite{bernstein1992} explicitly asked what the
phases of a matrix are and speculated the usage of the geometry of the numerical range of a matrix. Therein the authors also raised the question if there exists a different fundamental principle in MIMO control theory beyond the well-known small gain theorem by exploiting the phase information. However, decades later, a widely accepted phasic counterpart of the small gain theorem is still lacking in the literature. Not only that, a useful phase plot of a MIMO frequency response is nonexistent while the magnitude plot has been built into the computing environment MATLAB. In this sense, the current MIMO control theory is yet incomplete due to the splintering understandings of MIMO phases.

While MIMO phase maintains its mystery, there have been attempts over the years that reveal some useful clues. The early UK school of MIMO frequency domain theory \cite{postlethwaite1979complex} recognized the
importance of the characteristic phases of the loop transfer function but was not successful
in connecting them with the properties of the components in the feedback loop. Reference \cite{Macfarlane1981} defined the ``principal phases'' of a complex matrix as the arguments of the eigenvalues of the unitary part of its polar decomposition and formulated a small phase theorem based on this definition. However, the condition therein depends on both the phase and gain information, which somewhat deviates from the purpose of finding a phasic counterpart to the small gain theorem. References \cite{Chen,Looze,Anderson1988} made attempts to extend the Bode gain-phase integral relation to MIMO systems. Reference \cite{Owens} proposed the use of numerical range to incorporate both gain and phase information of perturbations in a bid to reduce the conservatism of the small gain theorem. Inspired by \cite{Owens}, the authors in \cite{Tits} defined the notion of median phase and phase spread by using the numerical range. Recently, reference \cite{LKDSM} used the phase spread as a performance measure of uncertain systems whose numerical ranges are in a cone sector. It is worth noting that the median phase and phase spread are in general insufficient to fully characterize the phases of MIMO systems. Just like an $m\times m$ LTI system has $m$ gains, i.e., $m$ singular values, at each frequency, one would naturally expect that it also has $m$ phases at each frequency. Exploitations of numerical ranges appeared in stability analysis of large-scale networks as well, for instance \cite{Lestas2012,Pates2017}.

Speaking of incorporating the phase information, one should also note the classic theory of positive real systems and recent studies of negative imaginary systems. Roughly speaking, we can think of SISO positive real systems as those whose phases lie within $[-\frac{\pi}{2},\frac{\pi}{2}]$ and SISO negative imaginary systems as those whose phases over the positive frequencies lie within $[-\pi,0]$. The phase theory developed in this paper will demonstrate that same thing can be said for MIMO systems. Research on positive real systems can be traced back to more than half a century ago and
has led to a rich theory through efforts of generations of researchers. See the books \cite{Anderson1973,BaoLee2007,Brogliato,DV1975} and the survey paper \cite{Kottenstette} for a review. Over the past two
decades, negative imaginary systems \cite{lanzon2008stability,petersen2010csm} and counter-clockwise dynamics \cite{Angeli2006} have attracted much attention. The abundant studies on these systems, concerning feedback stability, performance and beyond, give valuable insights in developing a general phase theory for MIMO LTI systems.

The main obstacle in enacting the phase concept for MIMO systems lies in the lack of a mathematical definition of matrix phases suited to control system theory. As control theorists started laying their eyes on the numerical range in describing the phase information of systems, mathematicians also turned their attention to the numerical range in their explorations of matrix phases \cite{DeprimaJohnson1974,FurtadoJohnson2001,FurtadoJohnson2003,ZhangFuzhen2015}. Nevertheless, it just appears that the understandings from the two communities had not got a chance to fully cross-pollinate.
Very recently, we initiated to adopt the canonical angles introduced in \cite{FurtadoJohnson2001} as the phases of a sectorial complex matrix whose numerical range does not contain the origin\cite{WCKQ2020}. We studied various properties of matrix phases, some of which are reviewed and extended later. This gives us a good starting point for conducting a systematic study of phase analysis and design for MIMO LTI systems.

The purpose of this paper is to present novel concepts and results that build foundations of a phase theory of MIMO LTI systems. We start with reviewing and extending the knowledge on matrix phases and then define phase responses of MIMO LTI systems based on matrix phases. The newly defined phase concept elevates the qualitative phasic descriptions including positive realness and negative imaginariness to a fully quantitative description. It leads to the birth of a small phase theorem, complementing the celebrated small gain theorem. The time-domain interpretation of the MIMO phase is also investigated. As for the computation, a sectored real lemma in parallel to the well-known bounded real lemma is obtained in terms of linear matrix inequalities (LMIs). We also extend our discussions to systems having poles and zeros on the imaginary axis. We absorb much nutrition from the existing studies on positive real systems, negative imaginary systems, (generalized) KYP lemma, dissipativity, and integral quadratic constraints (IQCs), etc. along the way.

The current paper exhibits significant original contributions compared to the authors' early conference paper \cite{CWKQ2019}. Firstly, phases were only defined for stable systems in \cite{CWKQ2019}, and are now extended to semi-stable systems with possible poles and zeros on the imaginary axis. Secondly, the small phase theorem is now formulated in a more general setting with its necessity carefully studied. Thirdly, a new induced phase interpretation is provided in the context of power signal spaces. In addition, a new sectored real lemma is devised which gives a necessary and sufficient condition for phase bounded systems while the conference version only gave a sufficient condition.

The remainder of the paper is organized as follows. Knowledge on matrix phases is reviewed and extended in Section~2. Phase responses of MIMO systems are defined in Section~3. A small phase theorem and its necessity are presented in Section~4. Time-domain interpretations of phase bounded systems are discussed in Section~5, followed by state-space conditions in Section~6. Extensions to systems having imaginary-axis poles and zeros are presented in Section~7. Section~8 concludes the paper. The notation in this paper is more or less standard and will be made clear as we proceed.

\section{Phases of a Complex Matrix}
A nonzero complex scalar $c$ can be represented in the polar form as $c=\sigma e^{i\phi}$ with $\sigma >0$ and $\phi$ taking values in a half open
$2\pi$-interval, typically $[0,2\pi)$ or $(-\pi,\pi]$.  Here $\sigma=|c|$ is called the modulus or the magnitude and $\phi=\angle c$ is called the
argument or the phase.  The polar form is particularly useful when two complex numbers are multiplied. It simply holds that $|ab| =|a| |b|$ and
$\angle (ab) = \angle a + \angle b \mbox{ mod $2\pi$}$.

It is well understood that an $n \times n$ complex matrix $C$ has $n$ magnitudes, served by the $n$ singular values
\[
\sigma (C) = \begin{bmatrix} \sigma_1 (C) & \sigma_2 (C) & \cdots & \sigma_n (C) \end{bmatrix}
\]
with
$\overline{\sigma}(C)\!=\!\sigma_1 (C)\! \geq \!\sigma_2 (C)\! \geq\! \cdots\! \geq\! \sigma_n(C)\! =\! \underline{\sigma}(C)$ \cite{HornJohnson}.
The magnitudes of a matrix possess a number of nice properties, among which the following inequalities
\begin{align}\label{gainmajorization}
\overline{\sigma}(AB)  \leq \overline{\sigma} (A) \overline{\sigma}(B),  \ \
\underline{\sigma}(AB)  \geq \underline{\sigma} (A) \underline{\sigma}(B)
\end{align}
are of particular interest to the control community.

In contrast to the magnitudes of a complex matrix $C$, how to define the phases of $C$ appears to be an unsettled issue.  An early attempt \cite{Macfarlane1981} defined the phases of
$C$ as the phases of the eigenvalues of the unitary part of its polar decomposition. This definition was motivated by the seeming generalization of the polar form of a scalar to the polar decomposition of a
matrix. However, phases defined this way do not have certain desired properties.

Recently, we discovered a more suitable definition of matrix phases based on numerical range \cite{WCKQ2020}. The numerical range, also called field of
values, of a matrix $C \!\in\! \mathbb{C}^{n\times n}$ is defined as
\[
W(C) = \{ x^*Cx: x \in \mathbb{C}^n \mbox{ with } \|x\|=1\},
\]
which, as a subset of
$\mathbb{C}$, is compact and convex, and contains the spectrum of $C$ \cite{horntopics}. Furthermore, the angular numerical range, also called angular field of values, of $C$ is defined as
\[
W'(C) = \{ x^*Cx: x \in \mathbb{C}^n, x\neq 0 \},
\]
which is the conic hull of $W(C)$ and is always a convex cone. The field angle of $C$, denoted by $\delta(C)$, is defined as the angle subtended by $W'(C)$ if $W'(C)$ is salient, i.e., does not contain a line through the origin, as $\pi$ if $W'(C)$ contains one line through the origin, and as $2\pi$ if $W'(C)$ is the whole complex plane. See \cite{horntopics} for more details.

If $0\!\notin\! W(C)$, then $W(C)$ is contained in an open half plane due to its convexity and $\delta(C)\!<\!\pi$. In this case, $C$ is said to be sectorial. It is known
that a sectorial $C$ is congruent to a diagonal unitary matrix that is unique up to a permutation \cite{Horn,ZhangFuzhen2015}, i.e., there exist a nonsingular $T$ and a diagonal unitary $D$ such that
$C\!=\!T^*DT$.
This factorization is called sectorial decomposition in \cite{ZhangFuzhen2015}. 
In such a factorization, the eigenvalues (i.e., the diagonal elements) of $D$ are distributed in an arc on the unit circle with length less than $\pi$.
We can then define the phases of $C$, denoted by $$\overline{\phi}(C)=\phi_1(C)\geq\phi_2(C)\geq\dots\geq\phi_n(C)=\underline{\phi}(C),$$ to be the phases of the eigenvalues of $D$ so that $\overline{\phi}(C)-\underline{\phi}(C) <\pi$. The phases of $C$ defined in this way are not uniquely determined,
but are rather determined modulo $2\pi$. If we make a selection of $\displaystyle \gamma(C) = [\overline{\phi}(C)+\underline{\phi}(C)]/2$, called the phase center of $C$, in $\mathbb{R}$, then the phases are uniquely determined. The phases are said to take the principal values if $\gamma (C)$ is selected in $(-\pi, \pi]$. The phases defined in this fashion resemble the canonical angles of $C$ introduced in \cite{FurtadoJohnson2001}. Let us denote
\[
\phi (C) = \begin{bmatrix} \phi_1 (C) & \phi_2 (C) & \cdots & \phi_n(C) \end{bmatrix}.
\]
However, we will not always select $\gamma(C)$ using its principal value. Following the standard way of selecting the phase of a complex scalar, we will select $\gamma(C)$ to make it continuous in the elements of $C$. In this way $\phi(C)$ will also be continuous in the elements of $C$.

\begin{figure}[htb]
\centering
\subfigure[A sectorial matrix]{
\begin{minipage}{0.22\textwidth}\centering
\includegraphics[scale=0.4]{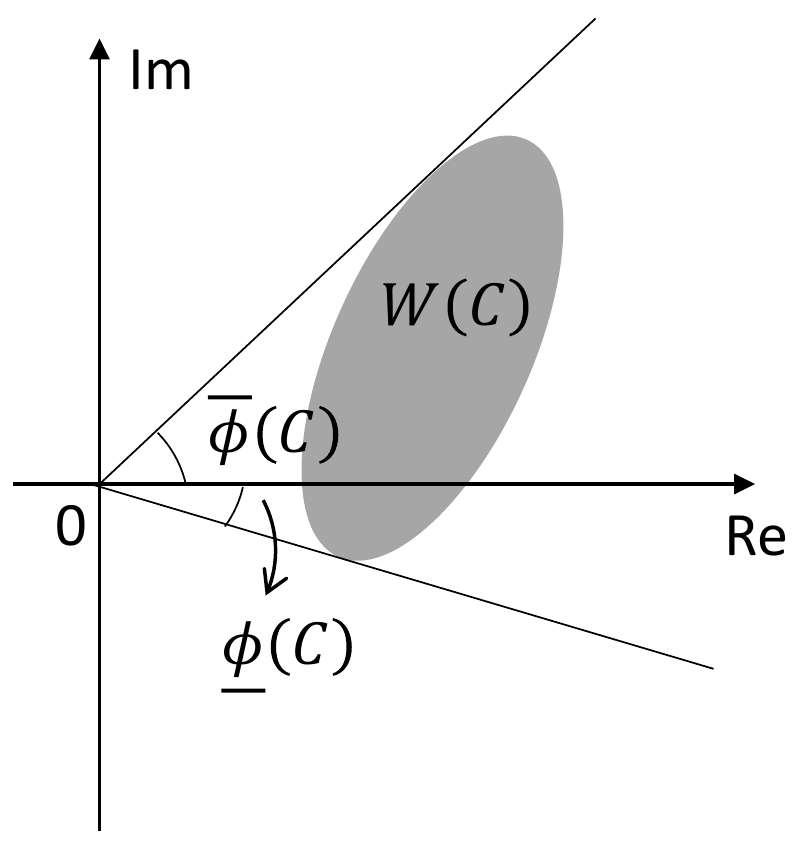}
\end{minipage}}
\hspace{20pt}
\subfigure[A quasi-sectorial matrix]{
\begin{minipage}{0.3\textwidth}\centering
\includegraphics[scale=0.4]{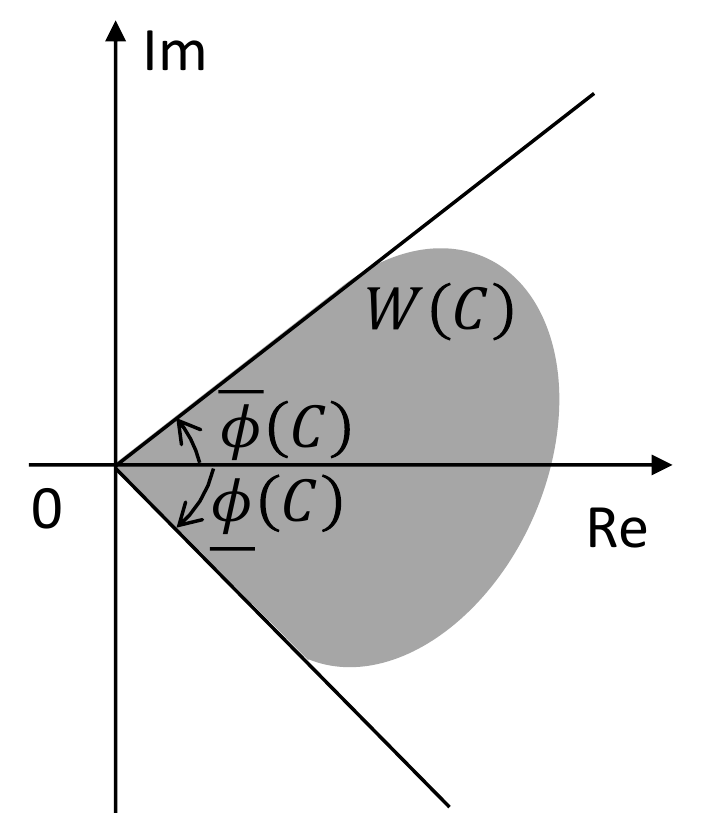}
\end{minipage}}
\subfigure[A semi-sectorial matrix]{
\begin{minipage}{0.3\textwidth}\centering
\includegraphics[scale=0.4]{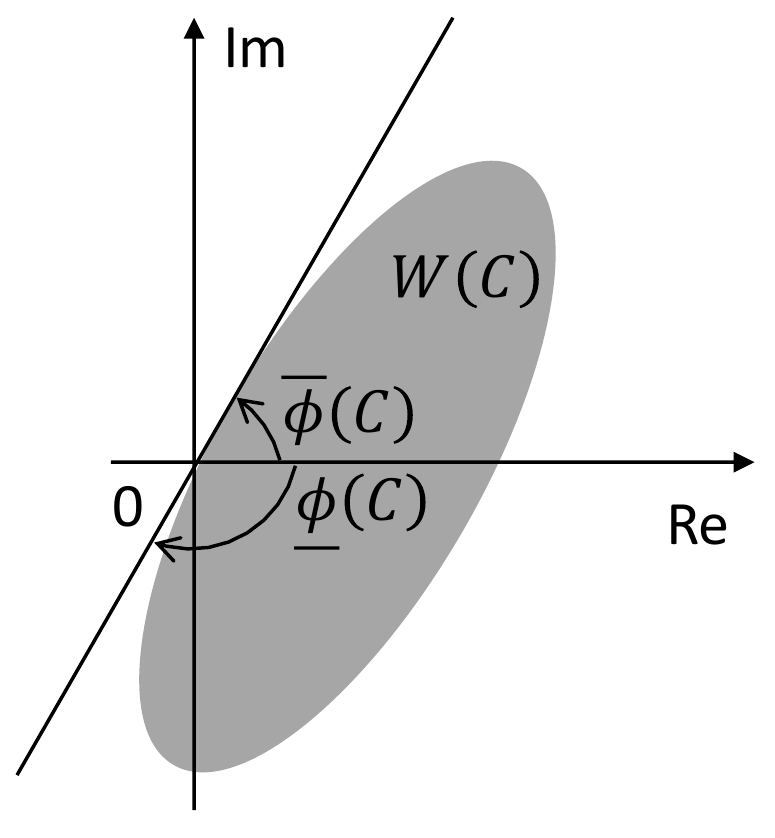}
\end{minipage}}
\vspace{-10pt}
\caption{Numerical ranges of typical sectorial, quasi-sectorial, and semi-sectorial matrices.}
\label{fig:semi-sectorial}
\end{figure}

A graphic interpretation of the phases is illustrated in \mbox{Fig. \ref{fig:semi-sectorial}(a)}. The two angles from the positive real axis to each of the two supporting rays of $W(C)$ are $\overline{\phi}(C)$ and $\underline{\phi}(C)$ respectively. The other phases of $C$ lie in between.

A non-sectorial matrix $C$ may also have $\delta(C) \!<\! \pi$. All positive semi-definite matrices fall into this case. An example of the numerical range of such a matrix is shown in Fig.~\ref{fig:semi-sectorial}(b). We see that 0 is a sharp point of the boundary of the numerical range. Such a matrix is said to be quasi-sectorial. Let $r=\mathrm{rank} (C)$. Then a quasi-sectorial $C$ has a decomposition
\begin{align}
C= U \begin{bmatrix} 0 & 0 \\ 0 & C_s \end{bmatrix} U^*\label{quasis}
\end{align}
where $U$ is unitary and $C_s \in \mathbb{C}^{r \times r}$ is sectorial \cite{FurtadoJohnson2003}, i.e., the range and kernel of $C$ are orthogonal and the compression of $C$ to its range is sectorial. Almost immediately we obtain the following useful characterization of quasi-sectorial matrices.

\begin{lemma}[\cite{Qiu2022}]\label{quasilemma}
A matrix $C$ is quasi-sectorial with phases taking principal values in $(-\frac{\pi}{2}+\alpha,\frac{\pi}{2}+\alpha)$, where $\alpha\in(-\pi,\pi]$, if and only if there exists $\epsilon>0$ such that
\begin{align}
e^{-j\alpha}C+e^{j\alpha}C^*\geq \epsilon C^*C.\label{quasiine}
\end{align}
\end{lemma}

The phases of $C$ are then defined as the phases of $C_s$.
Hence an $n \times n$ rank $r$ quasi-sectorial matrix $C$ has $r$ phases satisfying
$\overline{\phi}(C)=\phi_1(C)\geq\phi_2(C)\geq\dots\geq\phi_r(C)=\underline{\phi}(C)$.

Here comes a question. While an $n \times n$ zero matrix is clearly not sectorial, is it quasi-sectorial?
The answer is affirmative. It has zero number of phases and following conventions we have $\overline{\phi} (0)=-\infty$ and $\underline{\phi}(0)=\infty$.

A further larger class of matrices is the set of semi-sectorial matrices.
A matrix $C\in\mathbb{C}^{n\times n}$ is said to be semi-sectorial if $\delta(C)\leq \pi$, as illustrated in Fig.~\ref{fig:semi-sectorial}(c). A degenerate case of semi-sectorial matrices is when the numerical range has no interior and is given by a straight interval containing origin in its relative interior. In this case, $C$ is said to be rotated Hermitian and is subject to the following decomposition
\[
C= T^* \mathrm{diag} \{ 0_{n-r}, e^{j(\theta_0+\pi/2)} I,  e^{j(\theta_0-\pi/2)} I \} T.
\]
Here $\theta_0$, equal to the phase center $\gamma(C)$, is determined modulo $\pi$. It has two possible principal values in $(-\pi, \pi]$. The phases of $C$ are several copies of $\theta_0+\pi/2$
and several copies of $\theta_0-\pi/2$.

The generic case of semi-sectorial matrices is when the numerical range has nonempty interior. It is known \cite{FurtadoJohnson2003} that such a matrix has the following generalized sectorial decomposition
\begin{equation}
C=T^*\begin{bmatrix} 0_{n-r} & 0 & 0 \\ 0 & D & 0 \\  0 & 0 & E \end{bmatrix}T,
\label{gsd}
\end{equation}
where $
D=\mathrm{diag} \{ e^{j\theta_1}, \ldots , e^{j\theta_m}\}
$ with
$\theta_0+\pi/2 \geq \theta_1  \geq \cdots \geq \theta_m \geq \theta_0-\pi/2$ and
\[
E=\mathrm{diag} \left\{ e^{j\theta_0} \begin{bmatrix} 1 & 2 \\ 0 & 1 \end{bmatrix}, \dots , e^{j\theta_0} \begin{bmatrix} 1 & 2 \\ 0 & 1 \end{bmatrix} \right\}.
\]
In this case, the phases of $C$ are defined as $\theta_1, \ldots \theta_m$ and $(r-m)/2$ copies of
$\theta_0\pm\pi/2$. When $C$ is real, we often wish to have a real generalized sectorial decomposition, i.e., $T, D, E$ are real. Note that a real semi-sectorial $C$ has numerical range symmetric to the real axis and hence the phases are symmetric to zero, modulo $\pi$. The real generalized sectorial factorization of a real semi-sectorial $C$ can be established using a similar way as in \cite{FurtadoJohnson2003} and takes the form of (\ref{gsd}) with real nonsingular $T$,
\[
D=\mathrm{diag} \left\{ \begin{bmatrix} \cos \theta_1 & \sin \theta_1 \\ -\sin \theta_1 & \cos \theta_1 \end{bmatrix}, \ldots , \begin{bmatrix} \cos \theta_l & \sin \theta_l \\ -\sin \theta_l & \cos \theta_l \end{bmatrix}, \pm I \right\}
\]
where $l$ is a number less than $m/2$ and
\[
E= \pm \mathrm{diag} \left\{ \begin{bmatrix} 1 & 2 \\ 0 & 1 \end{bmatrix}, \dots ,  \begin{bmatrix} 1 & 2 \\ 0 & 1 \end{bmatrix} \right\}.
\]

From the above definitions, we can see that the phases of a semi-sectorial matrix are not continuous in its elements since the number of phases can change suddenly. However, a moment of thought suggests that the phases are continuous in the set of rank $r$ quasi-sectorial matrices and in the set of rank $r$ semi-sectorial matrices for each $0 \leq r \leq n$.

The notion of matrix phases subsumes the well-studied accretive and strictly accretive matrices \cite{Kato}, i.e., matrices with positive semi-definite and positive definite Hermitian parts respectively. In particular, the phases of a sectorial $C$ take principal values in $(- \pi/2, \pi/2)$ if and only if $C$ is strictly accretive; the phases of a semi-sectorial $C$ take principal values  in $[-\pi/2, \pi/2]$ if and only if $C$ is accretive. What is the role of quasi-sectorial matrices? A quasi-sectorial accretive matrix is called a quasi-strictly accretive matrix. A quasi-strictly accretive matrix cannot be identified from its Hermitian part. For example, $\begin{bmatrix} 1 & 0 \\ 0 & 0 \end{bmatrix}$ is quasi-strictly accretive while $\begin{bmatrix} 1 & 1 \\ -1 & 0 \end{bmatrix}$ is not though they have the same Hermitian part. However, according to Lemma \ref{quasilemma}, a matrix $C$ is quasi-strictly accretive if and only if $C^*+C \geq \epsilon C^*C$ for some $\epsilon >0$.

The matrix phases defined above have nice properties, of which comprehensive studies have been conducted in \cite{WCKQ2020} for sectorial matrices and in \cite{Qiu2022} for semi-sectorial matrices. The next few lemmas, whose proofs can be found in \cite{WCKQ2020,Qiu2022}, list some of the properties useful for our later developments. First, the phases of inverse $C^{-1}$ or even the Moore-Penrose generalized inverse $C^\dagger$ can be obtained from those of $C$ easily.

\begin{lemma}
Let $C\in \mathbb{C}^{n \times n}$ be semi-sectorial. Then
$\phi_i(C^\dagger)=-\phi_{\mathrm{rank} (C)-i+1} (C)$.
\end{lemma}

Next, the set of phase bounded semi-sectorial matrices
\begin{equation*}
\mathcal{C}[\alpha, \beta] =
\left\{C\in \mathbb{C}^{n\times n}: C \text{ is semi-sectorial and } [\underline{\phi}(C), \overline{\phi}(C)] \!\subset\! [\alpha, \beta] \right\}
\end{equation*}
where $0\leq \beta-\alpha<2\pi$, is a cone. 

\begin{lemma}
\label{convexcone}
If $\beta-\alpha\leq \pi$, then $\mathcal{C}[\alpha, \beta]$ is a convex cone.
\end{lemma}

Another important property pertinent to later developments in this paper concerns matrix product. In view of the magnitude counterpart in \eqref{gainmajorization}, one may expect
$\overline{\phi} (AB) \leq \overline{\phi}(A)+\overline{\phi}(B)$ to hold for semi-sectorial matrices $A$ and $B$. This, unfortunately, fails even for positive definite $A$ and
$B$. Notwithstanding, if we consider instead the eigenvalues of $AB$, the following weaker but useful result can be proved.

\begin{lemma} \label{thm: product_majorization} Let $A, B\in\mathbb{C}^{n\times n}$ be quasi-sectorial and semi-sectorial with phase centers $\gamma(A)$ and $\gamma(B)$ respectively. Then $AB$ has $\mathrm{rank} (AB)^2$ nonzero eigenvalues $\lambda_i(AB), i=1,\dots, \mathrm{rank}(AB)^2$, and $\angle \lambda_i (AB)$ can take values in $(\gamma(A)+\gamma(B)-\pi, \gamma(A)+\gamma(B)+\pi)$. Furthermore,
\begin{align}
\underline{\phi}(A) + \underline{\phi}(B) \leq \angle\lambda_i(AB) \leq \overline{\phi}(A) + \overline{\phi}(B).
\label{bound}
\end{align}
\end{lemma}

Inequalities in (\ref{bound}) underpin the development of a small phase theorem, much in the spirit of \eqref{gainmajorization} being the
foundation of the celebrated small gain theorem. To be more specific, recall that the singularity of matrix $I + AB$ plays an important role in
the stability analysis of feedback systems. It is straightforward to see that if $\overline{\sigma}(A) \overline{\sigma}(B)$ is less than the unity, then $I+AB$
is nonsingular. By contrast, one can show that if $\phi (A) + \phi (B)$ lies in $(-\pi, \pi)$, then $I+AB$ is nonsingular.

\begin{lemma}[Matrix small phase theorem] \label{matrix-spt}
Let $A \in \mathbb{C}^{n \times n}$ be quasi-sectorial with $\gamma (A) \in \mathbb{R}$. Then $\mathrm{det}\,(I + AB)\neq 0$ for all $B\in \mathcal{C}[\alpha, \beta]$ if and only if $[\alpha, \beta] \subset  (-\pi-\underline{\phi}(A), \pi- \overline{\phi}(A))$ modulo $2\pi$.
\end{lemma}

An immediate corollary is as follows: If $A$ is quasi-strictly accretive and $B$ is accretive, then $I+AB$ is nonsingular. This fact is probably the most fundamental form of the well-known passivity theorem in control theory.

\section{Phase Response of MIMO LTI Systems}
\label{sec:phase response}
Let $G$ be an $m\times m$ real rational proper stable transfer matrix, i.e., $G \in \mathcal{RH}_\infty^{m \times m}$. Then $\sigma(G(j\omega))$, the vector of singular values of $G(j \omega)$, is an $\mathbb{R}^m$-valued function of the frequency, which we call the magnitude response of $G$. The $\mathcal{H}_\infty$ norm of $G$, denoted by
$\|G\|_\infty=\sup_{\omega \in \mathbb{R}} \overline{\sigma}(G(j \omega))$, is of particular importance.




A system $G \in \mathcal{RH}_\infty^{m \times m}$ is said to be frequency-wise sectorial if $G(j \omega)$ is sectorial for all $\omega \in [-\infty,\infty]$.
For frequency-wise sectorial $G$, its DC phases are defined as $\phi(G(0))$. Since $G(0)$ is real, the DC phase center $\gamma(G(0))$ is either 0 or $\pi$, i.e., $G(0)$ is either strictly accretive or strictly anti-accretive. For the sake of simplicity, we assume throughout the paper that $\gamma(G(0))=0$. No generality is lost by this assumption since if $\gamma(G(0))=\pi$, we will then study $-G$ instead. Let $\phi(G(j\omega))$, the vector of phases of $G(j\omega)$, be defined so that
$\gamma(G(j\omega))$ is continuous in the frequency $\omega \in [-\infty,\infty]$. Then $\phi(G(j\omega))$ is an $\mathbb{R}^m$-valued continuous function of the frequency $\omega$,
which are called the phase response of $G$. The phase response of $G$ is an odd function of the frequency. Hence we will only focus on its half over the positive frequency.

For frequency-wise sectorial $G$, we define its maximum and minimum phases by
\begin{align*}
\overline{\phi}(G)=\sup_{\omega\in[0,\infty]}\overline{\phi}(G(j\omega)), \ \ 
\underline{\phi}(G)=\inf_{\omega\in[0,\infty]}\underline{\phi}(G(j\omega)),
\end{align*}
and its $\mathcal{H}_\infty$ phase sector, also called $\Phi_\infty$ sector, as
\begin{align*}
\Phi_\infty(G)= \displaystyle [ \underline{\phi}(G), \overline{\phi}(G) ],
\end{align*}
which serves as the counterpart to the $\mathcal{H}_\infty$ norm of $G$.
We also define the following cone of phase bounded frequency-wise sectorial systems
\begin{equation*}\label{pbs}
\mathfrak{C}(\alpha, \beta)=\{ G\in \mathcal{RH}_\infty^{m\times m}:\ G \text{ is frequency-wise sectorial and } \Phi_\infty(G)\subset (\alpha, \beta)\} .
\end{equation*}

Having defined the phase response of frequency-wise sectorial system $G$,
we can now plot $\sigma(G(j\omega))$ and $\phi(G(j\omega))$ together to complete the MIMO Bode plot of $G$,
laying the foundation of a complete MIMO frequency-domain analysis.

\begin{example}\label{example1}
The Bode plot of the system
\begin{align*}\footnotesize
G(s)\!=\!\frac{\begin{bmatrix}23s^3 \!+\! 17 s^2 \!+\! 29 s \!+\! 16 & -27 s^3 \!-\! 3 s^2 \!+\! 14 s \!+\! 14
  \\ -21 s^3 \!-\!  s^2 \!+\! 16 s \!+\! 14 &29 s^3 \!+\! 19 s^2 \!+\! 30 s \!+\! 16\end{bmatrix}}{4 s^3 + 5 s^2 + 2 s + 1}
\end{align*}
is shown in Fig. \ref{mimo-response}. One can see that $\Phi_\infty(G)\! \simeq\! [-135^\circ, 49^\circ]$.

\begin{figure}[htbp]
\centering
\includegraphics[scale=0.57]{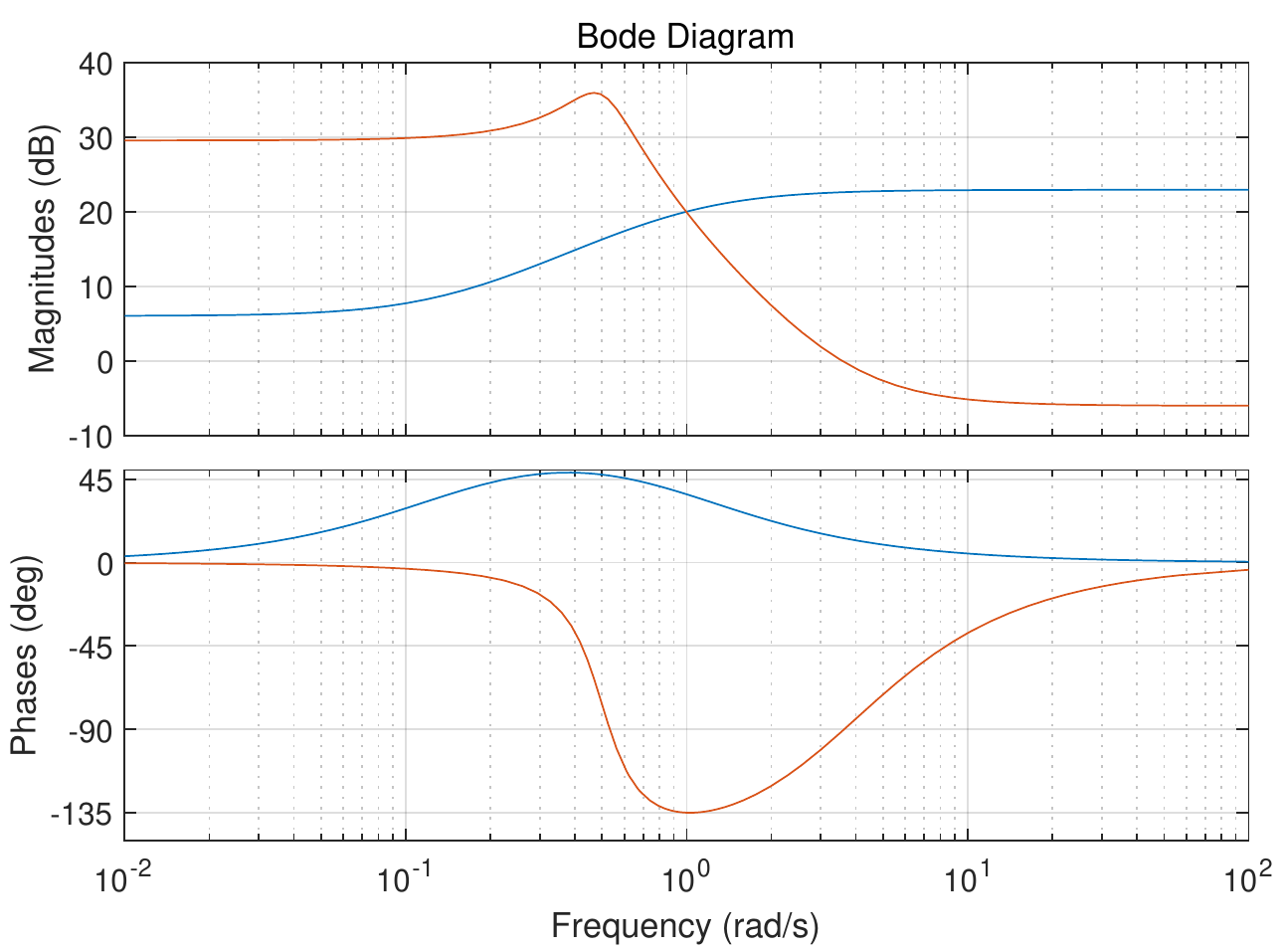}
\vspace{-10pt}
\caption{MIMO Bode plot of a frequency-wise sectorial system.}
\label{mimo-response}
\end{figure}
\end{example}

We say a $G$ is sectorial if it is in $\mathfrak{C}(\alpha, \beta)$ for some $\beta -\alpha \leq \pi$. Moreover, when $\beta -\alpha \leq \pi$, $\mathfrak{C}(\alpha, \beta)$ is a convex cone in view of \cite[Theorem 7.1]{WCKQ2020}.
Whether a system is sectorial or not can be read out from its phase plot over positive frequencies.
For instance, the system in Example \ref{example1} is not sectorial as its positive frequency phase response has a spread larger than $\pi$.

\begin{figure}[htb]
\centering
\subfigure[]{
\includegraphics[scale=0.75]{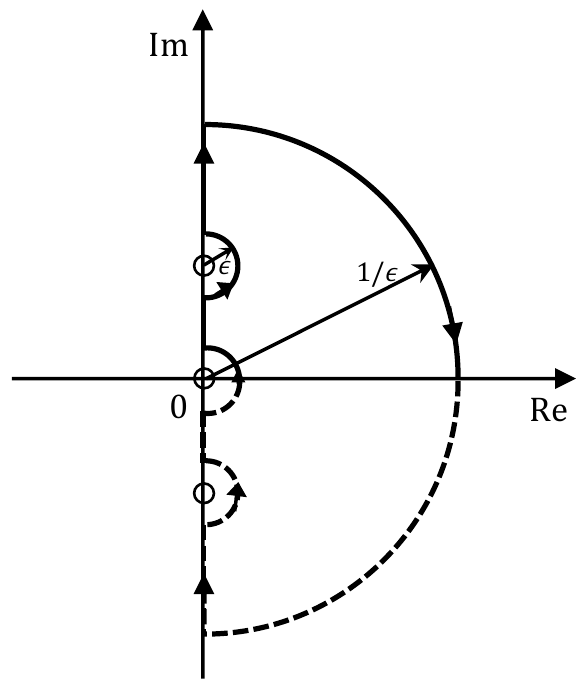}}
\hspace{5pt}
\subfigure[]{
\includegraphics[scale=0.57]{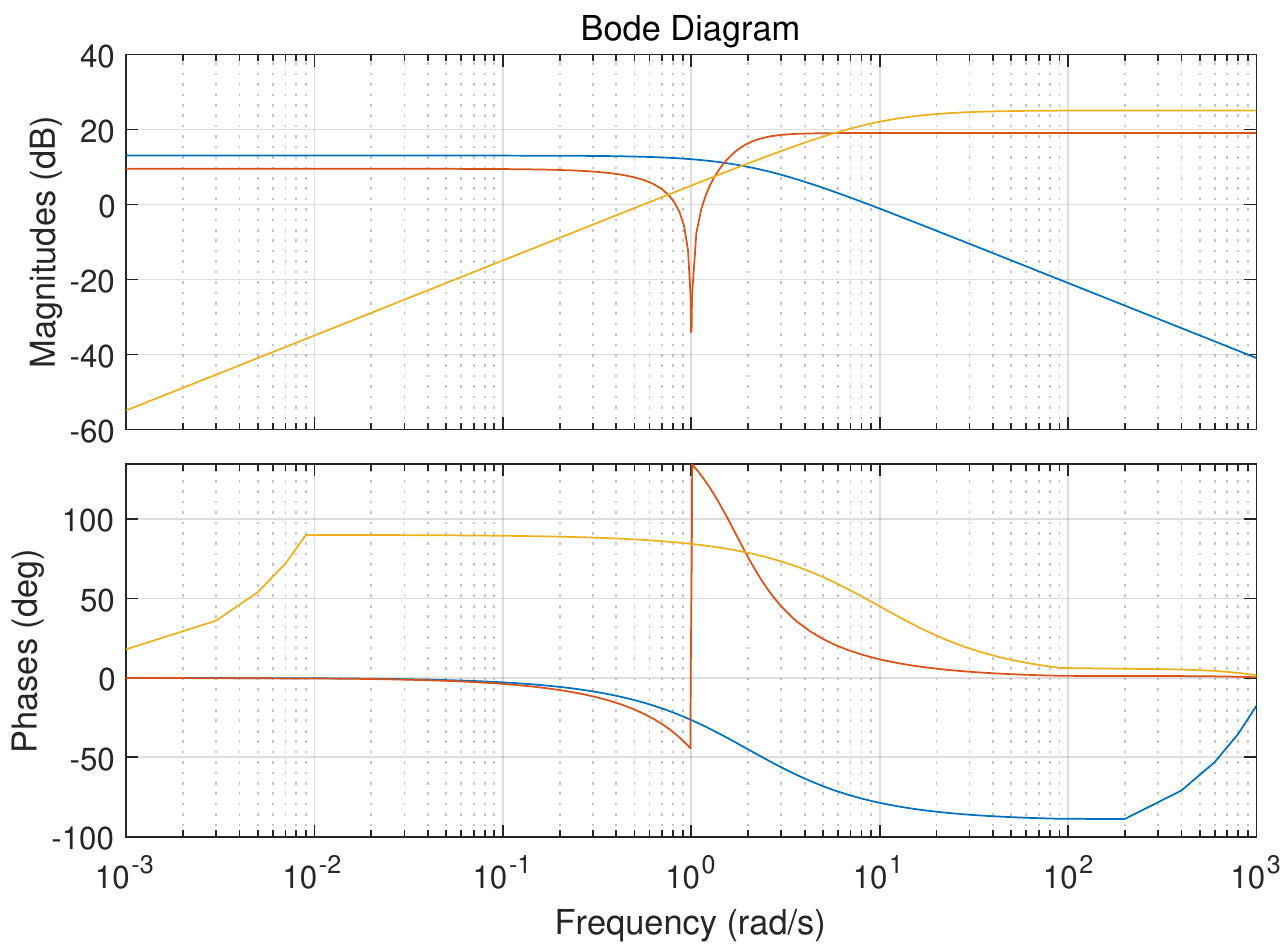}}
\vspace{-10pt}
\caption{(a) Indented $j\omega$-axis: ``o'' denotes the $j\omega$-axis zeros. (b) Bode plot of a $3\times 3$ frequency-wise semi-sectorial system with zeros $\{0, \pm j, \infty\}$.}
\label{fig:indent}
\vspace{-10pt}
\end{figure}

We wish to extend the systems under consideration to those with possible transmission zeros on the imaginary axis (including infinity), hence including strictly proper systems. We even wish to consider the systems with blocking zeros. Numerical issues may arise in such cases since the number of phases may drop at the zeros and the phases become discontinuous at these zeros. The discontinuity may cause difficulties in determining numerically the correct phase values around the discontinuous points, a difficulty that even exists in the current MATLAB implementation of the Bode diagram for SISO systems. We say $G \in \mathcal{RH}_\infty^{m \times m}$ is frequency-wise semi-sectorial if
\begin{enumerate}[1)]
    \item $G(j\omega)$ is semi-sectorial for all $\omega \in [-\infty,\infty]$; and
    \item there exists an $\epsilon^* > 0$ such that for all $\epsilon\leq \epsilon^*$, $G(s)$ has a constant rank and is semi-sectorial along the indented imaginary axis in Fig. \ref{fig:indent}(a), where the half-circle detours with radius $\epsilon$ are taken around the finite zeros of $G(s)$ on the imaginary axis and a half-circle detour with radius $1/\epsilon$ is taken if infinity is a zero of $G(s)$.
\end{enumerate}
Condition 1) implies condition 2) for SISO systems but not for general MIMO systems.
For frequency-wise semi-sectorial $G$, we also assume that the DC phase center $\gamma(G(0))$ (if $0$ is not a zero of $G$) or $\gamma(G(\epsilon))$ (if $0$ is a zero of $G$) is 0 and let $\gamma(G(s))$ be defined continuously along the indented imaginary axis.
With this construction, $\phi(G(s))$ is odd with respect to the indented imaginary axis. Hence we can focus on the part of the phase response corresponding to the upper half of the indented imaginary axis.



\begin{example}
The Bode plot of $G$ with transfer matrix given in \eqref{eq:ex2} is shown in Fig. \ref{fig:indent}(b). The system $G$ is frequency-wise semi-sectorial with zeros at $0, \pm j, \infty$. The phase response at extreme low frequency, extreme high frequency, and at frequency 1 clearly shows how the phases vary continuously along an indented imaginary axis.
\begin{equation}\footnotesize
\label{eq:ex2}
G(s)\!\!=\!\!\displaystyle\frac{\!\!\begin{bmatrix} 12 s^4 \!\!+\! 81 s^3 \!\!+\! 152 s^2 \!\!+\! 119 s \!\!+\! 110
& \!-6 s^4 \!-\! 6 s^3 \!+\! 10 s^2 \!+\! 22 s \!+\! 100 & -30 s^3 - 32 s^2 + 22 s - 20  \vspace{3pt} \\
-6 s^4 \!-\! 6 s^3 \!+\! 10 s^2 \!+\! 22 s \!+\! 100 & \!9 s^4 \!\!+\! 48 s^3 \!\!+\! 125 s^2 \!\!+\! 152 s \!\!+\! 140 & -6 s^4 \!-\! 36 s^3 \!-\! 22 s^2 \!+\! 44 s \!+\! 80  \vspace{3pt}\\
 -30 s^3 - 32 s^2 + 22 s - 20 &  -6 s^4 \!-\! 36 s^3 \!-\! 22 s^2 \!+\! 44 s \!+\! 80 & \!6 s^4 \!\!+\! 60 s^3 \!\!+\! 146 s^2 \!\!+\! 152 s \!\!+\! 200\!
\end{bmatrix}\!\!}{s^4 + 14 s^3 + 47 s^2 + 76 s + 60}\!.
\end{equation}
\end{example}

Let $G$ be frequency-wise semi-sectorial. Define the maximum and minimum phases of $G$ by
\begin{align*}
\overline{\phi}(G)=\sup_{\omega\in[0,\infty]}\overline{\phi}(G(j\omega)),\ \
\underline{\phi}(G)=\inf_{\omega\in[0,\infty]}\underline{\phi}(G(j\omega)),
\end{align*}
and define the $\Phi_\infty$ sector of $G$ as
$\Phi_\infty(G)= \displaystyle [ \underline{\phi}(G), \overline{\phi}(G) ]$.
Note that in the definition of $\overline{\phi}(G)$ and $\underline{\phi}(G)$, the sup and inf are taken in the non-negative frequency range, ignoring the indentation at the zeros of $G$. We also define the following closed cone of phase bounded frequency-wise semi-sectorial systems:
\begin{equation*}\label{pbs2}
\mathfrak{C}[\alpha, \beta]=\{G\in \mathcal{RH}_\infty^{m\times m}:\ G \text{ is frequency-wise semi-sectorial and } \Phi_\infty(G)\subset [\alpha, \beta]\}.
\end{equation*}
We say a system $G$ is semi-sectorial if it is in
$\mathfrak{C}[\alpha, \beta]$ for some $\beta-\alpha \leq \pi$. Moreover, when $\beta-\alpha\leq \pi$, $\mathfrak{C}[\alpha, \beta]$ is a convex cone in view of Lemma \ref{convexcone}.

A frequency-wise semi-sectorial system $G$ is said to be frequency-wise quasi-sectorial if $G(s)$ is quasi-sectorial for all $s$ on an indented imaginary axis. If in addition, $G\in\mathfrak{C}[\alpha, \beta]$ for some $\beta-\alpha < \pi$, then $G$ is said to be quasi-sectorial. The class of frequency-wise quasi-sectorial systems,  including all SISO stable systems which may have zeros on the imaginary axis, is of special interest.

The concept of phase unifies the well-known notions of positive real systems \cite{Anderson1973,Brogliato,Kottenstette}, negative imaginary
systems \cite{lanzon2008stability,petersen2010csm}, and relaxation systems \cite{Willems,Pates2019}.
A system $G\!\in\! \mathcal{RH}_\infty^{m \times m}$ is said to be positive real or passive if
$G(j\omega)+G^*(j\omega) \geq 0$ for all $\omega \in [0, \infty]$. It is said to be strongly positive real or input strictly passive if $G(j\omega)+G^*(j\omega) > 0$ for all $\omega \!\in\! [0, \infty]$. Positive real systems have the property that $G(s)\!+\!G^*(s) \!\geq\! 0$ for all $s$ with positive real parts. Hence stable positive real systems are those in $\mathfrak{C} [-\pi/2, \pi/2]$ and strongly positive real systems are those in $\mathfrak{C} (-\pi/2, \pi/2)$. A system in
$\mathfrak{C} [-\pi/2, \pi/2]$, i.e., a stable positive real system, is called output strictly passive if $G(j\omega)+G^*(j\omega) \geq \epsilon G^*(j\omega)G(j\omega)$ for some $\epsilon>0$ \cite{Kottenstette}. One can show that output strictly passive systems are exactly those in $\mathfrak{C} [-\pi/2, \pi/2]$ which are frequency-wise quasi-sectorial.

A system $G \in \mathcal{RH}_\infty^{m \times m}$ is said to be negative imaginary if
$(G(j\omega)-G^*(j\omega))/j \leq 0$ for all $\omega \in [0, \infty]$. It is easy to see that stable negative imaginary systems with $G(0)$ being accretive are those in $\mathfrak{C} [-\pi, 0]$.
A system $G \in \mathcal{RH}_\infty^{m \times m}$ is said to be a relaxation system if it can be written as $\displaystyle G(s) = G_0+ \sum_{i=1}^n \frac{G_i}{s+\lambda_i}$
where $G_i \geq 0$ for all $i=0, 1, \dots, n$ and $\lambda_i>0$ for all $i=1, \dots, n$.
It is easy to see that the relaxation systems are both
positive real and negative imaginary, and they belong to $\mathfrak{C}[-\pi/2, 0]$.
The class of frequency-wise semi-sectorial systems is much richer than positive real and negative imaginary systems. For example the system shown in Fig.~\ref{mimo-response} is neither positive real nor negative imaginary but it has a well-defined phase response.

Positive real, negative imaginary, and relaxation systems also cover certain unstable systems, namely those with poles possibly on the imaginary axis but not in the open right half plane. We will also extend our definition of frequency-wise semi-sectorial systems to this class of unstable systems. Things are more tedious in that case and so we leave them to Section~7.

\section{Small Phase Theorem}
\label{sec:spt}
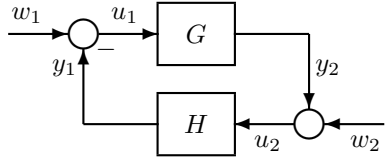
\begin{figure}[htbp]
\begin{center}
\begin{picture}(50,20)
\thicklines 
\put(0,17){\vector(1,0){8}} \put(10,17){\circle{4}}
\put(12,17){\vector(1,0){8}} \put(20,13){\framebox(10,8){$G$}}
\put(30,17){\line(1,0){10}} \put(40,17){\vector(0,-1){10}}
\put(38,5){\vector(-1,0){8}} \put(40,5){\circle{4}}
\put(50,5){\vector(-1,0){8}} \put(20,1){\framebox(10,8){$H$}}
\put(20,5){\line(-1,0){10}} \put(10,5){\vector(0,1){10}}
\put(5,10){\makebox(5,5){$y_1$}} \put(40,10){\makebox(5,5){$y_2$}}
\put(0,17){\makebox(5,5){$w_1$}} \put(45,0){\makebox(5,5){$w_2$}}
\put(13,17){\makebox(5,5){$u_1$}} \put(32,0){\makebox(5,5){$u_2$}}
\put(10,10){\makebox(6,10){$-$}}
\end{picture}
\vspace{-10pt}
\caption{A standard feedback system.}
\label{fdbk}
\end{center}
\end{figure}

Suppose $G$ and $H$ are $m\times m$ real rational proper transfer function matrices. The feedback interconnection of $G$ and $H$, as depicted in Fig. \ref{fdbk}, is said to
be stable if the Gang of Four matrix
\begin{align*}
G\#H=\begin{bmatrix}
  (I + HG)^{-1} & (I + HG)^{-1}H\\
  G(I + HG)^{-1} & G(I + HG)^{-1}H
\end{bmatrix}
\end{align*}
is stable, i.e., $G\#H \in \mathcal{RH}^{2m \times 2m}_\infty$.

The celebrated small gain theorem \cite{Zhou,LiuYao2016} is one of the most used results in robust control theory over the past half a century. A version of it states that for $G, H \in \mathcal{RH}^{m\times m}_\infty$, the feedback system $G\#H$ is stable if
\begin{align*}
\overline{\sigma}(G(j\omega)) \overline{\sigma}(H(j\omega))< 1
\end{align*}
for all $\omega \in [0, \infty]$.

There was an attempt to formulate a small phase theorem by using phases defined from the matrix polar decomposition \cite{Macfarlane1981}. However, the condition therein involves both phase and gain information and thus deviates from the initial purpose of having a phase counterpart of the small gain theorem.

Armed with the new definition of matrix
phases, we obtain the following theorem.

\begin{theorem}[Small phase theorem]\label{smallphase}
Let $G \in \mathcal{RH}^{m\times m}_\infty$ be frequency-wise quasi-sectorial  and $H\!\in\! \mathcal{RH}^{m\!\times\! m}_\infty$ be frequency-wise semi-sectorial. Then
$G\#H$ is stable if
\begin{align}\label{spi} 
\overline{\phi} (G(j\omega))+ \overline{\phi}(H(j\omega))  < \pi, \ \
\underline{\phi} (G(j\omega))+ \underline{\phi}(H(j\omega))  > -\pi 
\end{align}
for all $\omega \in [0,\infty]$.
\end{theorem}

\begin{proof}
Since $G,H\!\in\!\mathcal{RH}^{m\!\times\! m}_\infty$, by the generalized Nyquist criterion \cite{DesoerWang}, it suffices to show that $\det[I+G(j\omega)H(j\omega)] \!\neq\! 0$ for all $\omega\in[-\infty,\infty]$ and that the net number of encirclements of point $-1$ by the closed paths formed from the eigenloci of $G(s)H(s)$ along the imaginary axis is zero.

To this end, observe that when inequalities in (\ref{spi}) hold for all $\omega\in[0,\infty]$, due to the oddness of phase responses, they hold for all $\omega\!\in\![-\infty,\infty]$. Applying Lemma  \ref{thm: product_majorization}, we have
\begin{equation*}
\underline{\phi} (G(j\omega))+ \underline{\phi}(H(j\omega))\leq \angle\lambda_i(G(j\omega)H(j\omega)) \leq \overline{\phi} (G(j\omega))+ \overline{\phi}(H(j\omega)).
\end{equation*}
Hence, $-\pi<\angle\lambda_i(G(j\omega)H(j\omega))<\pi$ for all nonzero eigenvalues $\lambda_i(G(j\omega)H(j\omega))$
for all $\omega\in[-\infty,\infty]$. This implies that $\det[I+G(j\omega)H(j\omega)]\neq 0$ for all $\omega\in[-\infty,\infty]$. In addition, the encirclement of $-1$ by the eigenloci of $G(s)H(s)$ is excluded since there is no $\omega\!\in\![0,\infty]$ such that $\angle \lambda_i(G(j\omega)H(j\omega))\!=\!\pi$. This completes the proof.
\end{proof}


The small phase theorem generalizes a stronger version of the LTI passivity theorem \cite{DV1975}, which states that
$G\#H$ is stable if $G$ is output strictly passive and $H$ is stable passive. It also generalizes a stronger version of the negative imaginary theorem \cite{petersen2010csm}, which states that $G\#H$ is stable if $G$ is strictly proper strictly negative imaginary and $-H$ is stable negative imaginary.

Note that the small gain theorem provides a quantifiable tradeoff between the gains of $G$ and $H$, while the small phase theorem does the same but in terms of the phases of $G$ and $H$. In the literature, the notions of input feedforward passivity index and output feedback passivity index \cite{Vidyasagar1981,Wen1988,Kottenstette} have been used to characterize the tradeoff between
the surplus and deficit of passivity. It is our belief that the concept of MIMO system phases is more suited to this
task. Specifically, $\frac{\pi}{2}-\max\{|\underline{\phi}(G)|,|\overline{\phi}(G)|\}$ gives an angular measure of passivity, which we call the angular passivity index. The small phase theorem implies that if the sum of the angular passivity indexes of $G$ and $H$ are positive, then $G\#H$ is stable. In addition, one can see that $\pi-\max\{|\underline{\phi}(GH)|,|\overline{\phi}(GH)|\}$
yields a phase stability margin of $G\#H$.

It is known that the condition in the small gain theorem is necessary in the following sense \cite{Vidyasagar1985}. Let $r\in\mathcal{RH}_\infty$ be a scalar transfer function and define a ball of systems
$\mathfrak{B}(r) =\{H\!\in\!\mathcal{RH}^{m\times m}_\infty: \overline{\sigma}(H(j\omega))\!\leq\! |r(j\omega)|, \forall \omega \in [0,\infty]\}$.
Then, $G\#H$ is stable for all $H\in\mathfrak{B}(r)$ if and only if $\overline{\sigma}(G(j\omega)) < |r(j\omega)|^{-1}$ for all $\omega\in[0,\infty]$.

Analogously, let $h\in\mathcal{RH}_\infty$ be a scalar transfer function with $h^{-1}\in\mathcal{RH}_\infty$ and define a cone of frequency-wise semi-sectorial systems
\begin{multline*}
\mathfrak{C}(h)=\{H\in\mathcal{RH}^{m\times m}_\infty:\ \overline{\phi}(H(j\omega)) \leq \pi/2+\angle h(j\omega), \\
\underline\phi(H(j\omega)) \geq -\pi/2+\angle h(j\omega),
\forall \omega \in [0,\infty]\}.
\end{multline*}
This cone is just a set of weighted stable positive real systems. The phase uncertainty range is $\pi$ at each frequency. A special version of the small phase theorem containing a necessary and sufficient condition is as follows.

\begin{theorem}[Small phase theorem with necessity]
\label{them: spot with nece}
Let $G\in\mathcal{RH}^{m\times m}_\infty$ and $h\in\mathcal{RH}_\infty$ be a transfer function with $h^{-1} \in \mathcal{RH}_\infty$. Then, $G\#H$ is stable for all $H\in\mathfrak{C}(h)$ if and only if $G$ is frequency-wise quasi-sectorial and
\begin{align}\label{eq:sptn1} 
\overline{\phi}(G(j\omega)) < \pi/2-\angle h(j\omega), \ \
\underline\phi(G(j\omega)) > -\pi/2-\angle h(j\omega) 
\end{align}
for all $\omega\in [0,\infty]$.
\end{theorem}

\begin{proof}
The sufficiency follows directly from Theorem \ref{smallphase}. Suppose $G$ is not frequency-wise quasi-sectorial or one of the inequalities in (\ref{eq:sptn1}) is not satisfied. Then there exists a frequency $\omega_0 \in [0,\infty]$
such that $G(j\omega_0)$ is either not quasi-sectorial or one of the following holds
\begin{align*}
\overline{\phi}(G(j\omega_0))  \geq \pi/2-\angle h(j\omega_0), \ \
\underline\phi(G(j\omega_0))  \leq -\pi/2-\angle h(j\omega_0),
\end{align*}
which means that $A:=h(j\omega_0)G(j\omega_0)$ is not quasi-strictly accretive. The necessity follows if we can find $H \in \mathfrak{C}(h)$ such that $G\#H$ is unstable.

First we consider the case when $\omega_0=0$ or $\infty$. In this case, $A$ is real. There are two possibilities, one is that $A$ is completely non-accretive and the other is that $A$ is accretive but not quasi-strictly accretive. For the first possibility, let the smallest eigenvalue of $(A+A^*)/2$ be $\mu(A) < 0$. Notice that $\mu(A)I+(A-A^*)/2$ is nonsingular since the second term is skew-symmetric and hence has only imaginary eigenvalues. Set $B=-(\mu(A)I+(A-A^*)/2)^{-1}$, then $B$ is accretive and
\[
I+AB=(B^{-1}+A)B=(-\mu(A)I+(A+A^*)/2)B
\]
is singular. For the second possibility, it follows from the real generalized sectorial factorization that there is a real nonsingular matrix $T$ such that
\begin{align*}
A=T^* \mathrm{diag} \left\{ \begin{bmatrix} 0 & 1 \\ -1 & 0 \end{bmatrix}, \tilde{A}\right\} T
\text{ or }
A=T^* \mathrm{diag} \left\{ \begin{bmatrix} 1 & 2 \\ 0 & 1 \end{bmatrix}, \tilde{A} \right\} T
\end{align*}
where $\tilde{A}$ is accretive.
In both situations, set
$B=T^{-1} \mathrm{diag} \left\{\begin{bmatrix} 0 & 1 \\ -1 & 0 \end{bmatrix}, I\right\} T^{*-1}$.
Then again $B$ is accretive and $I+AB$ is singular.
The matrix $B$ obtained in this way is real, then $H(s)=h(s)B$ is in $\mathfrak{C}(h)$, and $G\# H$ is unstable.

Next we will consider the case when $\omega_0\in(0,\infty)$. In this case, $A$ is complex. Again there are two possibilities. From the same procedure as above, we can obtain complex accretive matrix $B$ such that $I+AB$ is singular. If we can find a positive real transfer matrix $\tilde{H} \in \mathcal{RH}^{m \times m}_\infty$ satisfying the interpolation condition
$\tilde{H}(j\omega_0)=B$,
then set $H=h\tilde{H}$. Consequently,
\begin{align*}
\overline{\phi}(H(j\omega)) &=\overline{\phi}(\tilde{H}(j\omega))+\angle h(j\omega)\leq \pi/2+\angle h(j\omega), \\
\underline{\phi}(H(j\omega)) &=\underline{\phi}(\tilde{H}(j\omega))+\angle h(j\omega)\geq -\pi/2+\angle h(j\omega)
\end{align*}
for all $\omega \in (0,\infty)$ and
$
I+G(j\omega_0) H(j\omega_0)
=I+h(j\omega_0)G(j\omega_0)h^{-1}(j\omega_0)H(j\omega_0)=I+AB
$
which is singular. Therefore, $H$ belongs to $\mathfrak{C}(h)$ and $G\#H$ is unstable.

To find such an $\tilde{H}$, we use Nevanlinna-Pick interpolation. First we find a positive real $\mathcal{H}_\infty^{m\times m}$ function $\tilde{H}_c$ such that
$\tilde{H}_c(j \omega_0)=B$
and
$\tilde{H}_c(-j \omega_0)=\overline{B}$.
This is a boundary Nevanlinna-Pick interpolation problem \cite{YoulaSaito1967} and it is solvable since both $B$ and $\overline{B}$ are accretive.
In general such an $\tilde{H}_c$ is complex rational. Let $\tilde{H}(s) \!=\! \frac{\tilde{H}_c(s) \!+\! \overline{\tilde{H}_c (\overline{s})}}{2}$. Then $\tilde{H}$
is a real rational positive real function satisfying $\tilde{H}(j \omega_0)\!=\!(B\!+\!\overline{\overline{B}})/2 \!=\! B$.
\end{proof}

The above proof critically depends on properties of positive real systems. The necessity of the positive real condition in the passivity theorem has been studied in \cite{KhongKao2019}.

Finally in this section, we examine how closed-loop system phases are connected to open-loop system phases. To this end, we name $S=(I+GH)^{-1}$ as the closed-loop sensitivity function. 

\begin{theorem}
Let $G\in \mathcal{RH}^{m\times m}_\infty$ be frequency-wise quasi-sectorial
and $H\in \mathcal{RH}^{m\times m}_\infty$ be frequency-wise semi-sectorial
and assume small-phase conditions (\ref{spi}) are satisfied. Then $SG$ and $HS$ are frequency-wise semi-sectorial and for each $\omega \in [0,\infty]$,
\begin{multline}
\left[ \underline{\phi}[S(j\omega)G(j\omega)],\ \overline{\phi}[S(j\omega) G(j\omega)] \right] \\
 \subset \left[
\min \left\{ \underline{\phi} [G(j\omega) ], - \overline{\phi} [H(j\omega)] \right\},  \max \left\{ \overline{\phi} [G(j\omega)], -\underline{\phi} [H(j\omega)] \right\} \right]
\label{inclusion1}
\end{multline}
and
\begin{multline}
\left[ \underline{\phi}[H(j\omega)S(j\omega)],\ \overline{\phi} [H(j\omega)S(j\omega)] \right] \\
 \subset \left[
\min \left\{-\overline{\phi} [G(j\omega)], \underline{\phi} [H(j\omega)] \right\},  \max \left\{ -\underline{\phi} [G(j\omega)], \overline{\phi} [H(j\omega)] \right\} \right]. \label{inclusion2}
\end{multline}
\end{theorem}

\begin{proof}
Since $G(j\omega)$ is quasi-sectorial, there is a unitary matrix $U$ so that
\[
G(j\omega)=U\begin{bmatrix} 0 & 0 \\ 0 & \tilde{G}(j\omega) \end{bmatrix} U^*
\]
where $\tilde{G}(j\omega)$ is $r \times r$ nonsingular and sectorial. Let
$
\begin{bmatrix} ? & ? \\ ? & \tilde{H} (j\omega) \end{bmatrix} = U^* H(j\omega) U
$
where $?$ represents entries that are irrelevant.
Then $\tilde{H}(j\omega)$ is also semi-sectorial and
\[
\left[\underline\phi[\tilde{H}(j\omega)], \overline\phi[\tilde{H}(j\omega)]\right] \subset [\underline\phi[H(j\omega)], \overline\phi[H(j\omega)]].
\]
It also holds
\[
S(j\omega)G(j\omega) = U\begin{bmatrix} 0 & 0 \\ 0 &  \left(I+\tilde{G}(j\omega) \tilde{H}(j\omega)\right)^{-1} \tilde{G}(j\omega) \end{bmatrix}U^*.
\]
Since $G(j\omega)$ and $H(j\omega)$ satisfy the small phase condition, $\tilde{G}(j\omega)$ and $\tilde{H}(j\omega)$ also satisfy the small phase condition. It then follows that $\phi_{r-i+1} (\tilde{G}^{-1}(j\omega))$ and $\phi_i (\tilde{H} (j\omega))$ must be contained simultaneously in a closed interval of length $\pi$. This means that $\tilde{G}^{-1} (j\omega) \!+\! \tilde{H}(j\omega)$ and its inverse, $\left(I+\tilde{G}(j\omega) \tilde{H}(j\omega)\right)^{-1} \tilde{G}(j\omega)$, are semi-sectorial.
By virtue of Lemma 1, we obtain that for $i=1, \dots, r$,
\begin{align*}
\phi_i[S(j\omega)G(j\omega)]
\! =\! -\! \phi_{r-i+1} \!\!\left[\tilde{G}^{-1} (j\omega) \!\!\left(I\!+\!\tilde{G}(j\omega)\tilde{H}(j\omega)\right) \right]\! 
\! =\! -\! \phi_{r-i+1}  \!\left[\tilde{G}^{-1} (j\omega) \!+\! \tilde{H}(j\omega) \right]\!.
\end{align*}
Also, denoting by $\mathrm{conv}\{\cdot\}$ the convex hull of a set, we have
\begin{align*}
\phi_i [ \tilde{G}^{-1}(j\omega) + \tilde{H}(j\omega)]
& \in \! \mathrm{conv} \! \left\{\! \left[\underline\phi[\tilde{G}^{-1} (j\omega)],
\overline\phi[\tilde{G}^{-1}(j\omega)]\right] \! \cup \! \left[\underline\phi[\tilde{H}(j\omega)], \overline\phi[\tilde{H}(j\omega)]\right] \!\right\} \\
& = \! \mathrm{conv} \! \left\{\! \left[-\overline\phi[\tilde{G}(j\omega)], -\underline\phi[\tilde{G}(j\omega)]\right] \! \cup \! \left[ \underline\phi[\tilde{H}(j\omega)], \overline\phi[\tilde{H}(j\omega)]\right] \!\right\} .
\end{align*}
It then follows that
\begin{multline*}
\left[\underline\phi[S(j\omega)G(j\omega)], \overline\phi[S(j\omega)G(j\omega)]\right]\\\subset \mathrm{conv} \left\{  \left[\underline\phi[G(j\omega)], \overline\phi[G(j\omega)]\right] \cup \left[ -\overline\phi[H(j\omega)], -\underline\phi[H(j\omega)] \right] \right\}
\end{multline*}
which is followed by (\ref{inclusion1}).

Inclusion (\ref{inclusion2}) can be proved in a similar way.
\end{proof}

By taking infima and suprema of the end points of various intervals in (\ref{inclusion1}) and (\ref{inclusion2}) over frequencies in $[0,\infty]$ and noting that both $\Phi_{\infty}(G)$ and $\Phi_{\infty}(H)$ contain $0$, we arrive at the following generalization of similar results in positive real and negative imaginary system analysis \cite{Kottenstette,lanzon2008stability,petersen2010csm}.

\begin{corollary}
Let $G, H \in \mathcal{RH}^{m\times m}_\infty$ be frequency-wise quasi-sectorial
and frequency-wise semi-sectorial respectively. Assume small-phase conditions (\ref{spi}) are satisfied. Then
$\Phi_\infty (SG) \!\subset \! \Phi_\infty(G) \cup (-\Phi_\infty(H))$ and
$\Phi_\infty (HS) \!\subset \! (-\Phi_\infty(G)) \cup \Phi_\infty(H)$.
\end{corollary}

\section{Time-domain Interpretation of MIMO Phase}
A system $G\in\mathcal{RH}_{\infty}^{m\times m}$ corresponds in the time-domain to a bounded linear time-invariant causal operator mapping an input signal space to an output signal space. An input-output gain is then induced on the system which measures the relative size of the output with respect to the input. When $\mathcal{L}_2$ signals or power signals  (representing transient or persistent signals respectively) are considered, the induced input-output gain is given by the $\mathcal{H}_{\infty}$ norm \cite{Zhou}. In this section, we explore an analogous time-domain interpretation of system phase. We first consider $\mathcal{L}_2$ signal space and then power signal space.

\subsection{Induced phase in $\mathcal{L}_2$ space}

We introduce some background on signal spaces and the Hilbert transform. The Hilbert transform of a signal $x(t)$ is defined as
\[
[\boldsymbol{H} x](t)= \frac{1}{\pi}\int_{-\infty}^\infty \frac{x(\tau)}{t - \tau} \, d\tau,
\]
provided that the above integral exists in the sense of Cauchy principal value. Denote by $\mathcal{L}_2^T(-\infty,\infty)$ the Hilbert space of complex-valued bilateral time functions. The Hilbert transform is then a unitary operator on $\mathcal{L}_2^T(-\infty,\infty)$. Let $\boldsymbol{F}$ be the usual Fourier transform on $\mathcal{L}^T_2(-\infty, \infty)$, i.e.,
$[\boldsymbol{F} x ] (j\omega) = \int_{-\infty}^{\infty} x(t) e^{-j\omega t} dt$.
Then
\begin{align}
[\boldsymbol{F}\boldsymbol{H}x](j\omega)=-j\sgn \omega [\boldsymbol{F}x](j\omega),\label{HTF}
\end{align}
where $\mathrm{sgn}$ is the sign function. One can see that the Hilbert transform introduces a phase shift of $-\frac{\pi}{2}$ on the positive frequency and a phase shift of $\frac{\pi}{2}$ on the negative frequency.
The Hilbert transform has been used extensively in signal processing and communication theory, especially in the time-frequency domain analysis \cite{HT}. The use of Hilbert transform in control is mostly known in Bode's gain-phase relationship and system identification, etc. We refer interested readers to \cite{Anderson1988,DFT,Chen} for more details.

The Fourier transform on $\mathcal{L}_2^T(-\infty,\infty)$ is an isometry onto $\mathcal{L}^\Omega_2(-\infty, \infty)$, the Hilbert space of complex-valued bilateral frequency functions with inner product
\begin{align*}
\langle u(j\omega),v(j\omega) \rangle=\frac{1}{2\pi}\int_{-\infty}^{\infty}u^*(j\omega)v(j\omega)d\omega
\end{align*}
for $u(j\omega),v(j\omega)\!\in\!\mathcal{L}^\Omega_2(\!-\infty, \!\infty)$. If we decompose $\mathcal{L}_2^\Omega (\!-\infty, \!\infty)$ into a positive frequency signal space and a negative frequency signal space as
\[
\mathcal{L}_2^\Omega(-\infty, \infty)= \mathcal{L}_2^\Omega(0, \infty) \oplus
 \mathcal{L}_2^\Omega(-\infty, 0),
\]
then this is clearly an orthogonal decomposition. This in turn leads to a natural orthogonal decomposition in $\mathcal{L}_2^T(-\infty, \infty)$
\[
\mathcal{L}_2^T(-\infty, \infty)= \boldsymbol{F}^{-1} \mathcal{L}_2^\Omega(0, \infty) \oplus
\boldsymbol{F}^{-1} \mathcal{L}_2^\Omega(-\infty, 0).
\]
It is known that the time functions in the first subspace can be analytically extended to the upper half complex plane \cite{king2009hilbert}, i.e., the first subspace is the $\mathcal{H}_2$ space in time-domain. We denote the first subspace by $\mathcal{H}_2^T$ and consequently the second subspace by $\mathcal{H}_2^{T^\perp}$.
Let $\boldsymbol{P}$ be the orthogonal projection onto
$\mathcal{L}^\Omega_2 (0,\infty)$ and $\boldsymbol{Q}$ be the orthogonal projection onto $\mathcal{H}_2^T$. In view of (\ref{HTF}), we have
\begin{align*}
\boldsymbol{Q}x=\frac{1}{2} (x+j \boldsymbol{H} x) \text{ and }
(I-\boldsymbol{Q})x=\frac{1}{2} (x-j \boldsymbol{H} x),
\end{align*}
the analytic part (containing only positive frequency content) and the skew-analytic part (containing only negative frequency content) of $x$ respectively. In addition, we have
$\|\boldsymbol{Q}x\|_2=\|(I-\boldsymbol{Q})x\|_2=\frac{1}{\sqrt{2}}\|x\|_2$.

A full picture of the relationships among these signal spaces is shown in the commutative diagram as in Fig. \ref{commutativediagram}. Interested readers can compare this diagram with the one usually seen in the literature \cite{FrancisHinfinity,Zhou} which depicts the relationships between time-domain spaces $\mathcal{L}_2^T(-\infty,\infty),\mathcal{L}_2^T(0,\infty),\mathcal{L}_2^T(-\infty,0)$ and frequency-domain spaces $\mathcal{L}_2^{\Omega}(-\infty,\infty),\mathcal{H}_2,\mathcal{H}_2^{\perp}$.

\setlength{\unitlength}{0.75mm}
\begin{figure}[htb]
\begin{center}
\begin{picture}(68,62)
\thicklines
\put(38,58){\vector(-1,0){10}}
\put(28,60){\vector(1,0){10}}
\put(53,41){\makebox(10,10){$\boldsymbol{P}$}}
\put(-3,41){\makebox(10,10){$\displaystyle \frac{I+j\boldsymbol{H}}{2}$}}
\put(54,38){\vector(0,1){16}}
\put(49,54){\makebox(10,10){$\mathcal{L}^\Omega_2(0, \infty)$}}
\put(28,58){\makebox(10,10){$\boldsymbol{F}$}}
\put(28,50){\makebox(10,10){$\boldsymbol{F}^{-1}$}}
\put(7,54){\makebox(10,10){$\mathcal{H}_2^T$}}
\put(12,38){\vector(0,1){16}}
\put(7,28){\makebox(10,10){$\mathcal{L}^T_2(-\infty, \infty)$}}
\put(12,28){\vector(0,-1){16}}
\put(-3,15){\makebox(10,10){$\displaystyle \frac{I-j\boldsymbol{H}}{2}$}}
\put(7,2){\makebox(10,10){$\mathcal{H}_2^{T^\perp}$}}
\put(38,32){\vector(-1,0){10}}
\put(28,34){\vector(1,0){10}}
\put(28,32){\makebox(10,10){$\boldsymbol{F}$}}
\put(28,24){\makebox(10,10){$\boldsymbol{F}^{-1}$}}
\put(38,6){\vector(-1,0){10}}
\put(28,8){\vector(1,0){10}}
\put(28,6){\makebox(10,10){$\boldsymbol{F}$}}
\put(28,-2){\makebox(10,10){$\boldsymbol{F}^{-1}$}}
\put(49,28){\makebox(10,10){$\mathcal{L}^\Omega_2(-\infty,\infty)$}}
\put(54,28){\vector(0,-1){16}}
\put(56,15){\makebox(10,10){$I-\boldsymbol{P}$}}
\put(49,2){\makebox(10,10){$\mathcal{L}^\Omega_2(-\infty, 0)$}}
\end{picture}
\end{center}
\vspace{-10pt}
\caption{A commutative diagram.}
\label{commutativediagram}
\end{figure}

Now let $\mathcal{L}_2^{\mathbb{R}}(\!-\infty,\infty)$ be the real Hilbert space of real-valued signals in $\mathcal{L}_2^T(-\infty,\infty)$. Let $\boldsymbol{G}: \mathcal{L}^{\mathbb{R}}_2(-\infty, \infty)\!\rightarrow\!\mathcal{L}^{\mathbb{R}}_2(-\infty, \infty)$ be the linear operator corresponding to $G(s)\in\mathcal{RH}_\infty$. Both $\mathcal{H}_2^T$ and $\mathcal{H}_2^{T^\perp}$ are invariant subspaces of $\boldsymbol{G}$. We define the positive frequency numerical range and the negative frequency numerical range of $\boldsymbol{G}$ as
\begin{align*}
W_+(\boldsymbol{G})&:= \!\{\langle \boldsymbol{Q}u, \boldsymbol{G}u \rangle:\ u \!\in \mathcal{L}_2^{\mathbb{R}} (-\infty, \infty),\|u\|_2=1\}\\
W_-(\boldsymbol{G})&:= \!\{\langle (I\!-\!\boldsymbol{Q})u, \boldsymbol{G}u \rangle:\ u \!\in \mathcal{L}_2^{\mathbb{R}} (-\infty, \infty),\|u\|_2=1\}
\end{align*}
respectively. Note that
$\overline{\langle \boldsymbol{Q}u, \boldsymbol{G}u \rangle}=\langle (I-\boldsymbol{Q})u, \boldsymbol{G}u \rangle$
for all $u\in\mathcal{L}_2^{\mathbb{R}}(-\infty,\infty)$. This implies that $W_+(\boldsymbol{G})$ and $W_-(\boldsymbol{G})$ are symmetric with respect to the real axis. Denote the closure of $W_+(\boldsymbol{G})$ by $\mathrm{cl}\left\{ W_+(\boldsymbol{G})\right\}$.

\begin{proposition}\label{L2analysis}
Let $G\in\mathcal{RH}_{\infty}^{m\times m}$ be frequency-wise semi-sectorial. It holds
\begin{align*}
\mathrm{cl}\left\{W_+(\boldsymbol{G})\right\}\!=\! \mathrm{cl}\; \mathrm{conv}\left\{\frac{1}{2}W(G(j\omega)): \omega\!\in\![0,\infty]\right\}.
\end{align*}
\end{proposition}

\begin{proof}
We shall show the following inclusions
\begin{align}
    &\mathrm{cl}\left\{W_+(\boldsymbol{G})\right\}\subset\mathrm{cl}\; \mathrm{conv}\left\{\frac{1}{2}W(G(j\omega)): \omega\in[0,\infty]\right\}, \label{inclus1}\\
    &\mathrm{cl}\; \mathrm{conv}\left\{\frac{1}{2}W(G(j\omega)): \omega\in[0,\infty]\right\}\subset\mathrm{cl}\left\{W_+(\boldsymbol{G})\right\}.\label{inclus2}
\end{align}

First, for any $u\!\in\!\mathcal{L}_2^{\mathbb{R}}(-\infty,\infty)$ such that $\|u\|_2=1$, we have
\begin{align*}
\langle \boldsymbol{Q}u, \boldsymbol{G}u \rangle=\int_{-\infty}^{\infty}[\boldsymbol{Q}u]^*(t)[\boldsymbol{G}u](t) dt
=\frac{1}{2\pi}\int_{0}^{\infty} [\boldsymbol{F} u]^*(j\omega) G(j\omega)[\boldsymbol{F} u](j\omega)d\omega
\end{align*}
and
$ \|\boldsymbol{Q}u\|_2^2=\frac{1}{2}$. It follows that $\langle \boldsymbol{Q}u, \boldsymbol{G}u \rangle\!\in\! \mathrm{cl}\; \mathrm{conv}\left\{\frac{1}{2}W(G(j\omega)): \omega\in[0,\infty]\right\}$ and thus proves the inclusion (\ref{inclus1}).

Next we show the inclusion (\ref{inclus2}). For arbitrary $\omega_0 \in (0,\infty)$ and $x\!\in\!\mathbb{C}^m,\|x\|_2\!=\!1$, one can construct $u\!\in\! \mathcal{L}_2^{\mathbb{R}}(-\infty,\infty)$ so that
\[
[\boldsymbol{F}u](j\omega) = \begin{cases} \sqrt{\frac{\pi}{2\epsilon}}x & \text{if } |\omega - \omega_0| < \epsilon  \\\sqrt{\frac{\pi}{2\epsilon}}x^*& \text{if }|\omega+\omega_0|<\epsilon \\0 & \text{otherwise}, \end{cases}
\]
where $\epsilon > 0$. Clearly, $\|u\|_2 = 1$. Moreover, we have
\begin{align*}
\langle \boldsymbol{Q}u, \boldsymbol{G} u \rangle = \frac{1}{2\pi}\int_0^\infty [\boldsymbol{F}u]^*(j\omega) G(j\omega) [\boldsymbol{F}u](j\omega) d\omega
\end{align*}
which approaches to $\frac{x^*G(j\omega_0)x}{2}$ as $\epsilon \!\to\! 0$. This implies that $\frac{x^*G(j\omega_0)x}{2} \!\in\! \mathrm{cl}\left\{  W_+(\boldsymbol{G})\right\}$. One can verify that such an inclusion also holds for $\omega_0=0$ with a slight modification of the construction of $u$. Hence, $\frac{1}{2}W(G(j\omega)) \subset \mathrm{cl}\left\{W_+(\boldsymbol{G})\right\}$ for all $\omega \in [0,\infty)$. Since $\mathrm{cl}\left\{W_+(\boldsymbol{G})\right\}$ is closed and convex, the inclusion (\ref{inclus2}) follows.
\end{proof}

Note that Proposition \ref{L2analysis} has an important implication for semi-sectorial systems. Specifically, for a semi-sectorial system $G$, by definition, $\overline{\phi}(G)-\underline{\phi}(G)\leq\pi$. The above proposition then implies that $W_+(\boldsymbol{G})$ has two supporting rays from the origin which subtend an angle no greater than $\pi$. This further gives $\Phi_{\infty}(G)$ an ``induced phase'' type of interpretation. See the following theorem. The proof follows directly from Proposition \ref{L2analysis} and is thus omitted.

\begin{theorem}
Let $G\in\mathcal{RH}_{\infty}^{m\times m}$ be semi-sectorial. Then,
\begin{align*}
\underline{\phi}(G)=\inf_{z \in W_+(\boldsymbol{G})} \angle z \text{ and }
\overline{\phi}(G)=\sup_{z \in W_+(\boldsymbol{G})} \angle z.
\end{align*}
\end{theorem}

\subsection{Induced phase in power signal space}
Very often, it is also important to understand how a system behaves under persistent external disturbances or command signals. To this end, we consider the real signal space
\[\mathcal{P}=\mathcal{L}_2^{\mathbb{R}}\oplus \{u(t): u(t) \text{ is real finite periodic}\},
\]
which consists of responses of semi-stable LTI systems (the order of any pole on the imaginary axis of every element is at most one) when given $\mathcal{L}_2$ inputs. Although the discussions to follow can be extended to more general power signal space, we comment that $\mathcal{P}$ is representative enough to bring out the main points while maintaining technical simplicity.


For a signal $u(t)\in\mathcal{P}$, its auto-correlation matrix is
\begin{align*}
R_{uu}(\tau)=\lim_{T\rightarrow \infty}\frac{1}{2T}\int_{-T}^T u(t+\tau)u(t)'dt.
\end{align*}
The power of $u(t)$ is given by
\begin{align*}
\|u\|_{\mathcal{P}}=\sqrt{\mathrm{Tr} [R_{uu}(0)]}=\sqrt{\lim_{T\rightarrow \infty}\frac{1}{2T}\int_{-T}^T \|u(t)\|^2 dt},
\end{align*}
which defines a semi-norm on $\mathcal{P}$, where $\mathrm{Tr}$ is an abbreviation of trace. Taking Fourier transform of $R_{uu}(\tau)$ gives rise to the power spectral density matrix of $u(t)$
\begin{align*}
S_{uu}(j\omega)=\int_{-\infty}^{\infty}R_{uu}(\tau)e^{-j\omega\tau}d\tau .
\end{align*}
By Parseval's identity, we have
$\|u\|_{\mathcal{P}}=\sqrt{\frac{1}{2\pi}\int_{-\infty}^{\infty}\mathrm{Tr}[S_{uu}(j\omega)]d\omega}$.

Now, let $u(t),y(t)\in\mathcal{P}$. The cross-correlation matrix between $u(t)$ and $y(t)$ is
\begin{align*}
R_{uy}(\tau)=\lim_{T\rightarrow \infty}\frac{1}{2T}\int_{-T}^T u(t+\tau)y(t)'dt.
\end{align*}
Taking the Fourier transform of $R_{uy}(\tau)$ gives the cross power spectral density matrix
\begin{align*}
S_{uy}(j\omega)=\int_{-\infty}^{\infty}R_{uy}(\tau)e^{-j\omega\tau}d\tau.
\end{align*}
We define a semi-definite inner product $\langle \cdot,\cdot\rangle$ on $\mathcal{P}$:
\begin{align*}
\langle u,y\rangle =\lim_{T\rightarrow \infty}\frac{1}{2T}\int_{-T}^T u(t)'y(t)dt.
\end{align*}
An adaption of Parseval's identity tells that $\langle u,y\rangle$ can also be computed from the cross power spectral density $S_{uy}(j\omega)$, i.e.,
\begin{align}
\langle u,y\rangle =\frac{1}{2\pi}\int_{-\infty}^{\infty}\mathrm{Tr}[S_{yu}(j\omega)]d\omega.\label{cpsd}
\end{align}
For more details on power signal analysis, we refer interested readers to \cite{Zhou1994partI, champeney1987handbook}.

We now return to the theme of this subsection. It is known that the phase of a SISO system has a nice time-domain interpretation via electrical circuit. Given a stable scalar transfer function $g(s)$, one can associate with it the admittance function of a one-port circuit. When a sinusoidal voltage $u(t)=\cos \omega t$ is applied on the circuit, the output current is also sinusoidal
$y(t)=|g(j\omega)|\cos(\omega t+\angle g(j\omega))$,
where $\angle g(j\omega)$ represents the phase shift induced by the circuit at frequency $\omega$. If we define the real power $P$, reactive power $Q$, and complex power $S=P+jQ$ in the standard way \cite{arthur2000power}, then simple calculations yield that $\angle g(j\omega)=\angle S$. Note that $\cos \angle S$ gives the power factor which measures the trade-off between real and reactive power.

When it comes to MIMO systems, the phase shifts between input and output signals can no longer be found directly from the phase responses as for SISO systems. Nevertheless, we can still utilize the notion of real power and reactive power to describe the phase shifts induced by the system.
Given $G\in\mathcal{RH}_{\infty}^{m\times m}$, we associate with it the admittance matrix function of an $m$-port circuit. Let an input voltage $u(t)\!\in\!\mathcal{P}$ be such that $\|u(t)\|_{\mathcal{P}}\!=\!1$. The Hilbert transform of $u(t)$ exists and is also in $\mathcal{P}$ \cite{champeney1987handbook}. Let the corresponding output current be $y(t)\in\mathcal{P}$. According to circuit theory \cite{EP}, the real power $P$ and reactive power $Q$ produced by the circuit are given by
\begin{align*}
P=\langle u,y \rangle \text{ and }
Q=-\langle \boldsymbol{H}u,y\rangle,
\end{align*}
respectively.
The complex power is then given by
$S=P+jQ$.
As in the SISO case, $\angle S$ encodes the trade-off between real power and reactive power and is considered as a measure of the phase shift between $u$ and $y$.

In view of (\ref{cpsd}),
\begin{align*}
P=\langle u,y\rangle=\frac{1}{2\pi}\int_{-\infty}^{\infty}\mathrm{Tr}[S_{yu}(j\omega)]d\omega.
\end{align*}
By using the frequency-domain characteristics of Hilbert transform, one can show
\begin{align*}
Q=-\langle \boldsymbol{H}u,y\rangle=\frac{1}{2\pi}\int_{-\infty}^{\infty}-j\sgn \omega \cdot\mathrm{Tr}[S_{yu}(j\omega)]d\omega.
\end{align*}
It follows that
\begin{align}
S=P+jQ=\frac{1}{\pi}\int_{0}^{\infty}\mathrm{Tr}[S_{yu}(j\omega)]d\omega.\label{S}
\end{align}
For an LTI system $G(s)\in\mathcal{RH}_{\infty}^{m\times m}$ with input $u$ and output $y$, where $u,y\in\mathcal{P}$, it holds
$S_{yu}(j\omega)=G(j\omega)S_{uu}(j\omega)$.
Substituting this into (\ref{S}) yields
\begin{align}
S=\frac{1}{\pi}\int_{0}^{\infty}\mathrm{Tr}[G(j\omega)S_{uu}(j\omega)]d\omega.\label{Snum}
\end{align}

Let us first consider the case where the input is a sinusoidal voltage
\[u(t)=a\cos \omega_0(t)+b\sin \omega_0(t),\]
where $a,b\in\mathbb{R}^m$, $\left\|\begin{bmatrix}a'&b'\end{bmatrix}'\right\|_2\!=\!\sqrt{2}$. Applying equation (\ref{Snum}) yields
$S=d^*G(j\omega_0)d$,
where $d=\frac{a-jb}{\sqrt{2}}$. Moreover, one can show that
$\left\{S: \left\|\begin{bmatrix}a'&b'\end{bmatrix}'\right\|_2\!=\!\sqrt{2}\right\}\!=\!W(G(j\omega_0))$.
This is an important observation as it indicates that the phase shift produced by a frequency-wise semi-sectorial $G$ with a sinusoidal input at frequency $w_0$ is fully determined by $W(G(j\omega_0))$. Unlike the SISO case, the phase shift of a MIMO LTI system is direction dependent. Specifically, $\angle S=\angle d^*G(j\omega_0)d$ describes the phase shift produced by $G(s)$ at frequency $\omega_0$ under direction $d$. Depending on the direction, the phase shift varies in the interval $[\underline{\phi}(G(j\omega_0)),\overline{\phi}(G(j\omega_0))]$.

Having understood the phase shift induced by the system under sinusoidal inputs, now we turn our attention to general signals in $\mathcal{P}$. It turns out that in this case, for a frequency-wise semi-sectorial $G\in\mathcal{RH}_{\infty}^{m\times m}$, there holds
\begin{align}
\left\{S:\|u\|_{\mathcal{P}}=1\right\}=\mathrm{conv}\{W(G(j\omega)): \omega\in[0,\infty]\},\label{complexpowerset}
\end{align}
which can be shown in analogy to Proposition \ref{L2analysis}. It leads to an ``induced phase'' interpretation for the $\mathcal{H}_{\infty}$ phase sector of a semi-sectorial system when persistent input signals are allowed. See the following theorem. The proof is straightforward from identity (\ref{complexpowerset}) and is omitted for brevity.
\begin{theorem}
Let $G\in\mathcal{RH}_{\infty}^{m\times m}$ be semi-sectorial. Then,
\begin{align*}
\underline{\phi}(G)=\inf_{\|u\|_{\mathcal{P}}=1}\angle S \text{ and }
\overline{\phi}(G)=\sup_{\|u\|_{\mathcal{P}}=1}\angle S.
\end{align*}
\end{theorem}
One can see that only the numerical ranges over the positive frequency correspond to possible phase shifts, while the entire negative spectrum is redundant. This is sensible if one recalls Fourier's original definition of the Fourier transform which is based on real basis functions and positive frequency \cite{Fourier}.

\section{Sectored Real Lemma}
State-space methods have unique advantages in computing $\mathcal{H}_{\infty}$ norm (bounded real lemma) and verifying positive realness (positive real lemma) \cite{LiuYao2016}. Consider $G\!\in\!\mathcal{RH}_\infty^{m\times m}$ with a minimal realization $\left[\begin{array}{c|c}A & B \\ \hline C & D\end{array} \right]$. The bounded real lemma says $\|G\|_{\infty}\!<\!\gamma$ if and only if there is $X\!>\!0$ satisfying the
LMI
\begin{align*}
\begin{bmatrix}A'X+XA&XB\\B'X&0\end{bmatrix}+\begin{bmatrix}C'C&C'D\\D'C&D'D-\gamma^2 I\end{bmatrix}<0.
\end{align*}

In this section, we derive a sectored real lemma in parallel with the bouned real lemma, which gives an LMI condition for verifying $\Phi_{\infty}(G)\subset (\alpha,\beta)$. Then $\Phi_{\infty}(G)$ can be computed via a bisection search. The result is extended to semi-sectorial systems with possible poles on the imaginary axis in the next section.


Note that $\Phi_{\infty}(G)\subset (\alpha,\beta)$ is a frequency-domain description over only the positive frequency range. This makes the classical KYP lemma, which builds the equivalence between frequency-domain inequalities over the whole frequency range and LMIs no longer applicable. To overcome this difficulty, we exploit the more recent generalized KYP lemma \cite{IwasakiHara2005}, which handles inequalities over partial frequency ranges.

The generalized KYP lemma \cite{IwasakiHara2005} connects inequalities on curves in the complex plane with LMIs. Consider the curves characterized by the set
\begin{align*}
\mathrm{\mathbf{\Lambda}}(\Sigma,\Psi)=\left\{\lambda\in\mathbb{C}\left|\begin{bmatrix}\lambda\\1\end{bmatrix}^*\!\Sigma\begin{bmatrix}\lambda\\1\end{bmatrix}=0,\begin{bmatrix}\lambda\\1\end{bmatrix}^*\!\Psi\begin{bmatrix}\lambda\\1\end{bmatrix}\geq0\right.\right\},
\end{align*}
where $\Sigma,\Psi$ are given $2\times 2$ Hermitian matrices. 
Denote by $\otimes$ the Kronecker product of matrices. A version of the generalized KYP lemma is as follows.

\begin{lemma}[Generalized KYP lemma \cite{IwasakiHara2005}]\label{lem: gKYP}
\mbox{Let $A\in\mathbb{C}^{n\times n}$}, $B\!\in\!\mathbb{C}^{n\times m}$, $M\!=\!M^*\!\in\!\mathbb{C}^{(n+m)\times(n+m)}$, and $\mathrm{\mathbf{\Lambda}}(\Sigma,\Psi)$ be curves in the complex plane. Suppose $[A|B]$ is controllable. Let $\mathrm{\mathbf{\Omega}}$ be the set of eigenvalues of $A$ in $\mathrm{\mathbf{\Lambda}}(\Sigma,\Psi)$. Then
\begin{align}
\begin{bmatrix}(\lambda I-A)^{-1}B\\I\end{bmatrix}^*M\begin{bmatrix}(\lambda I-A)^{-1}B\\I\end{bmatrix}\leq 0\label{FDI}
\end{align}
for all $\lambda\in \mathrm{\mathbf{\Lambda}}(\Sigma,\Psi)\backslash \mathrm{\mathbf{\Omega}}$ if and only if there exist Hermitian $X$ and $Y$ such that
\begin{align*}
Y\geq0,\quad \begin{bmatrix}A&B\\I&0\end{bmatrix}^*(\Sigma\otimes X+\Psi\otimes Y)\begin{bmatrix}A&B\\I&0\end{bmatrix}+M\leq 0.
\end{align*}
\end{lemma}

\vspace{6pt}


Of particular interest is when $\Sigma\!=\!\begin{bmatrix}0&1\\1&0\end{bmatrix}$ and $\Psi\!=\!\begin{bmatrix}0&j\\-j&0\end{bmatrix}$. In this case, $\mathrm{\mathbf{\Lambda}}(\Sigma,\Psi)$ is the positive imaginary axis corresponding to the positive frequency range $\omega\in[0,\infty]$.

Now, let $G\in\mathcal{RH}_{\infty}^{m\times m}$ be quasi-sectorial. Observe that $\Phi_{\infty}(G)\!\subset(\alpha,\beta)$, where $0< \beta-\alpha\leq \pi$, if and only if $G(j\omega)$ has phases lie in $(-\frac{\pi}{2}+\eta_1,\frac{\pi}{2}+\eta_1)\cap(-\frac{\pi}{2}+\eta_2,\frac{\pi}{2}+\eta_2)$ for all $\omega\in[0,\infty]$, where $\eta_1=\alpha+\frac{\pi}{2}$ and $\eta_2=\beta-\frac{\pi}{2}$. By \mbox{Lemma \ref{quasilemma}}, this holds if and only if there exists $\epsilon_i>0,i=1,2$ such that
\begin{align}
    e^{-j\eta_i} G(j\omega)+e^{j\eta_i} G^*(j\omega)\geq \epsilon_i G^*(j\omega)G(j\omega), \;i=1,2,\label{hsfdi1}
\end{align}
for all $\omega\!\in\![0,\infty]$.
The above inequalities can be rewritten into
\begin{align}
 \begin{bmatrix}(j\omega I-A)^{-1}B\\I\end{bmatrix}^*\!M_i\!\begin{bmatrix}(j\omega I-A)^{-1}B\\I\end{bmatrix}\leq 0,\;i=1,2,\label{hsfdi}
\end{align}
where
\begin{align}
M_i\!=\!\begin{bmatrix}\!\epsilon_iC'C\!&\!-e^{j\eta_i}C'\!+\!\epsilon_i C'D\!\\\!-e^{-j\eta_i}C\!+\!\epsilon_i D'C\!&\!\!-e^{-j\eta_i}D\!-\!e^{j\eta_i}D'\!+\!\epsilon_i D'D\!\!\end{bmatrix}\!.\label{Mi}
\end{align}
With these preparations, we are ready to state the sectored real lemma.

\begin{theorem}[Sectored real lemma]\label{thm: half-bound}
Let $G\in\mathcal{RH}_{\infty}^{m\times m}$ be quasi-sectorial with a minimal realization $\left[\begin{array}{c|c}\!A\! & \!B\! \\ \hline \!C\! & \!D\!\end{array} \right]$. Then $\Phi_\infty(G)\subset (\alpha, \beta)$, $0<\beta-\alpha\leq \pi$, if and only if there exist $\epsilon_i>0$ and Hermitian matrices $X_i, Y_i$, $i=1,2$, such that
\begin{align*}
Y_i\geq0, \begin{bmatrix}A&B\\I&0\end{bmatrix}'\!\begin{bmatrix}0&X_i+jY_i\\X_i-jY_i&0\end{bmatrix}\!\begin{bmatrix}A&B\\I&0\end{bmatrix}+M_i\leq 0,
\end{align*}
where $M_i$ are given by (\ref{Mi}).
\end{theorem}

\begin{proof}
From the above analysis, we know that $\Phi_\infty(G) \subset (\alpha, \beta)$ if and only if the inequalities (\ref{hsfdi}) hold for all $\omega\!\in\![0,\infty]$. 
Applying the generalized KYP lemma with $\Sigma\!=\!\begin{bmatrix}0&1\\1&0\end{bmatrix}$ and $\Psi=\begin{bmatrix}0&j\\-j&0\end{bmatrix}$ yields that the inequalities (\ref{hsfdi}) hold for all $\omega\in[0,\infty]$ if and only if there exist $\epsilon_i>0$ and Hermitian matrices $X_i,Y_i$, $i=1,2$ satisfying the LMIs in the theorem.
\end{proof}

When $\alpha=-\frac{\pi}{2}$ and $\beta=\frac{\pi}{2}$, we have $\eta_1=\eta_2=0$. Then, the inequalities (\ref{hsfdi1}) reduce to
    $G(j\omega)+G^*(j\omega)\geq \epsilon G^*(j\omega)G(j\omega)$
for some $\epsilon>0$ for all $\omega\in[0,\infty]$, which is the frequency-domain characterization of output strictly passive systems. One can easily verify that in this case, the two LMIs in Theorem \ref{thm: half-bound} reduce to one LMI
\begin{align*}
\begin{bmatrix}A'X\!+\!XA&XB\\B'X&0\end{bmatrix}\!+\!\begin{bmatrix}\epsilon C'C&-C'\!+\!\epsilon C'D\\-C\!+\!\epsilon D'C&-D\!-\!D'\!+\!\epsilon D'D\end{bmatrix}\!\leq \!0,
\end{align*}
which is consistent with the one in \cite[Lemma 2]{Kottenstette} for output strictly passive systems.

\section{Extensions to Semi-stable Systems}\label{sec: extension}

Now we extend the main concepts and results of phase theory to semi-stable systems, i.e., systems that may have poles on the imaginary axis but no poles in the open right half plane.

\subsection{Phase response of semi-stable systems}

Let $G$ be an $m\times m$ real rational proper semi-stable system with $j\Omega$ being the set of poles on the imaginary axis.
We say $G$ is frequency-wise semi-sectorial if
\begin{enumerate}[1)]
    \item $G(j\omega)$ is semi-sectorial for all $\omega \in [-\infty,\infty]\backslash \Omega$; and
    \item there exists an $\epsilon^* > 0$ such that for all $\epsilon\leq \epsilon^*$, $G(s)$ has a constant rank and is semi-sectorial along the indented imaginary axis, where the half-circle detours with radius $\epsilon$ are taken around both the poles and finite zeros of $G(s)$ on the frequency axis and a half-circle detour with radius $1/\epsilon$ is taken if infinity is a zero of $G(s)$.
\end{enumerate}


For a semi-stable frequency-wise semi-sectorial
$G$, we also assume $\gamma(G(0))$ (if 0 is neither a pole nor a zero) or $\gamma(G(\epsilon))$
(if 0 is a pole or zero) is $0$ and let $\gamma(G(s))$ be defined continuously on the indented imaginary
axis. A typical example of frequency-wise semi-sectorial
systems is $\frac{A}{s^n}$, where $A$ is an accretive matrix. The phases of this system at all positive finite frequencies are $\phi_i(A)-n\frac{\pi}{2}$, $i=1, \ldots, \rank (A)$.

\begin{figure}[htb]
    \centering
    \includegraphics[scale=0.57]{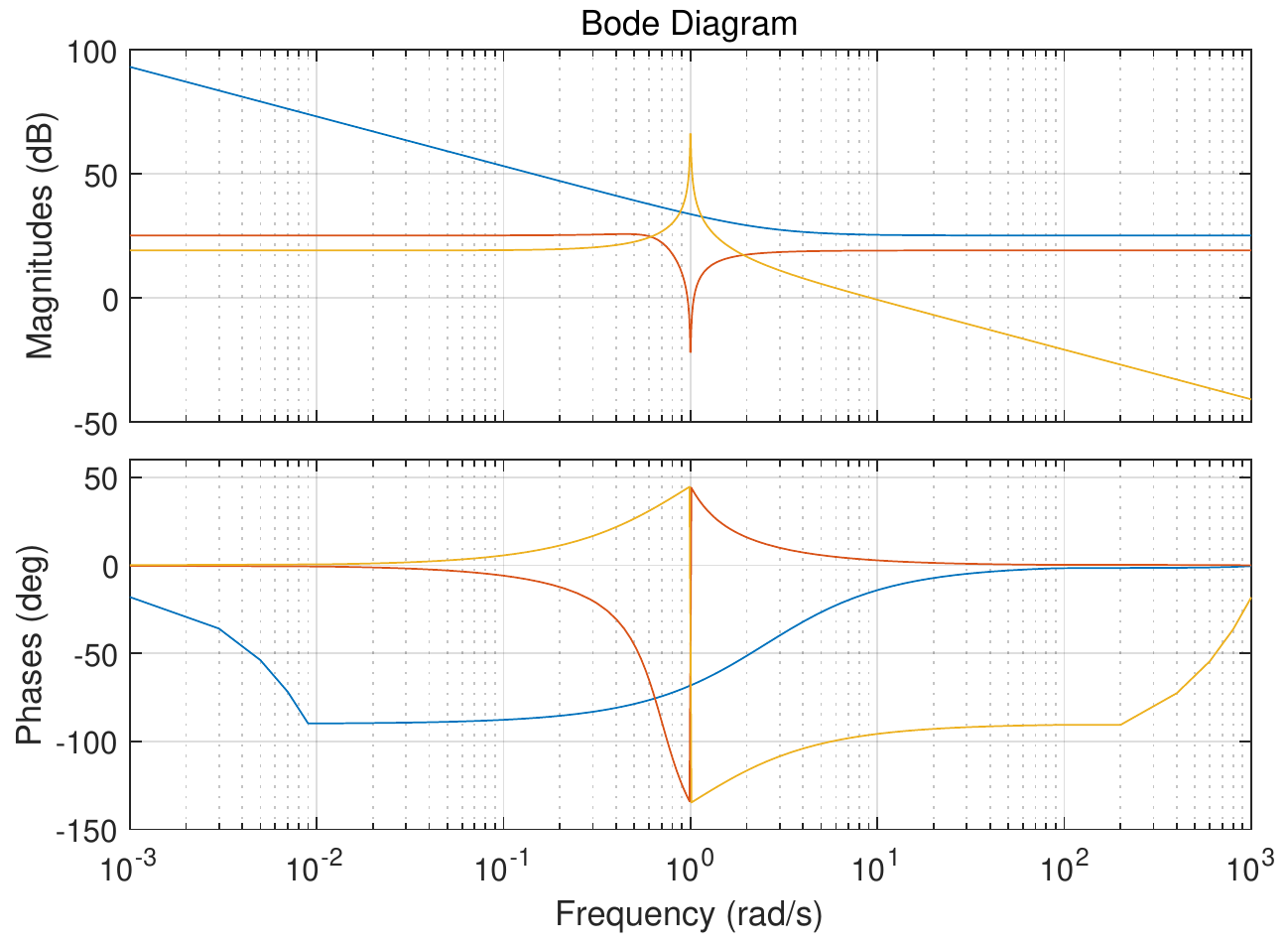}
    \vspace{-10pt}
    \caption{Frequency response of a $3\times 3$ semi-stable frequency-wise semi-sectorial system.}
    \label{fig:Bode_polezero}
\end{figure}

\begin{example}
The Bode plot of a semi-stable frequency-wise semi-sectorial system $G$ with transfer matrix \eqref{eq:ex3} is shown in Fig.~\ref{fig:Bode_polezero}. The system $G$ has poles at $0, \pm j$ and zeros at $\pm j, \infty$.
\begin{equation}\scriptsize
\label{eq:ex3}
\begin{split}
&G(s)=\\
&\displaystyle \frac{\!\!\begin{bmatrix} 12 s^5 \!\!+\! 20 s^4 \!\!+\! 39 s^3 \!\!+\! 25 s^2 \!\!+\! 19 s \!+\! 5   &    12 s^5 \!\!+\! 16 s^4 \!\!+\! 18 s^3 \!\!+\! 26 s^2 \!\!+\! 14 s \!+\!10    &   -28 s^4 - 12 s^3 - 38 s^2 - 8 s - 10 \\
\!12 s^5 \!\!+\! 16 s^4 \!\!+\! 18 s^3 \!\!+\! 26 s^2 \!\!+\! 14 s \!+\! 10  & 18 s^5 \!\!+\! 56 s^4 \!\!+\! 60 s^3 \!\!+\! 76 s^2 \!\!+\! 34 s \!+\! 20  &  \!\!-12 s^5 \!\!-\! 44 s^4 \!\!-\! 30 s^3 \!\!-\! 64 s^2 \!\!-\! 22 s \!-\! 20 \\
-28 s^4 - 12 s^3 - 38 s^2 - 8 s - 10  & \!\!-\!12 s^5 \!\!-\! 44 s^4 \!\!-\! 30 s^3 \!\!-\! 64 s^2 \!\!-\! 22 s \!\!-\! 20  &\! 24 s^5 \!\!+\! 50 s^4 \!\!+\! 63 s^3 \!\!+\! 70 s^2 \!\!+\! 37 s \!+\! 20
\end{bmatrix}\!\!}{ 2 s^5 + s^4 + 3 s^3 + s^2 + s}.
\end{split}
\end{equation}
\vspace{-10pt}
\end{example}

For semi-stable frequency-wise semi-sectorial $G$, we also define its maximum and minimum phases by $\overline{\phi}(G)=\sup_{\omega\in[0,\infty]\backslash\Omega}\overline{\phi}(G(j\omega)),
\underline{\phi}(G)=\inf_{\omega\in[0,\infty]\backslash\Omega}\underline{\phi}(G(j\omega))$
and its $\Phi_\infty$ sector by
$\Phi_\infty(G)\!=\! \displaystyle [ \underline{\phi}(G), \overline{\phi}(G)]$.
In this section, without ambiguity we let
\begin{align*}
\mathfrak{C}[\alpha, \beta]=\{G \text{ is semi-stable:}\ G \text{ is frequency-wise semi-sectorial and } \Phi_\infty(G)\subset [\alpha, \beta]\}.
\end{align*}
Then, $G$ is said to be semi-sectorial if it is in $\mathfrak{C}[\alpha, \beta]$ for some $\beta-\alpha\leq \pi$.

The phase concept subsumes the notion of positive real systems. A real rational proper semi-stable system $G$ is said to be positive real if
\begin{enumerate}[1)]
\item $G(j\omega)+G^*(j\omega)\geq 0$ for all $\omega \in [-\infty,\infty]\backslash \Omega$; and
\item if $j\omega_0$ is a pole of any element of $G$, then it is a simple pole and the residue matrix $\lim_{s\rightarrow j\omega_0}(s-j\omega_0)G(s)$ is positive semi-definite.
\end{enumerate}
Hence, positive real systems are those in $\mathfrak{C}[-\pi/2, \pi/2]$.

\subsection{Generalized small phase theorem}

The following result gives a phase condition for the feedback stability of a semi-stable system and a stable system.

\begin{theorem}[Generalized small phase theorem]
\label{thm:gspt}
Let $G$ be semi-stable frequency-wise semi-sectorial with $j\Omega$ being the set of poles on the imaginary axis and $H\in\mathcal{RH}_\infty$ be frequency-wise sectorial. Then $G\#H$ is stable if
\begin{align}\label{eq: gspt1}
\sup_{w\in[0,\infty]\backslash \Omega}\left[\;\overline{\phi}(G(j\omega))\!+\!\overline\phi(H(j\omega))\right]\!<\!\pi,
\inf_{w\in[0,\infty]\backslash \Omega}\left[\;\underline{\phi}(G(j\omega))\!+\!\underline\phi(H(j\omega))\right]\!>\!-\pi.
\end{align}
\end{theorem}

\begin{proof}
We again utilize the generalized Nyquist criterion. When inequalities in (\ref{eq: gspt1}) hold, by Lemma  \ref{thm: product_majorization} and the oddness of the phase responses, we have
\begin{align}
    -\pi<\angle\lambda_i(G(s)H(s))<\pi\label{eigine}
\end{align}
for all nonzero eigenvalues $\lambda_i(G(s)H(s))$ and for all $s$ in the indented imaginary axis.

Let $\omega_0\in\Omega$.
Suppose that $j\omega_0$ is a pole of order $k$ of an element of $G$. Following a similar argument to the proof in \cite[Theorem 2.7.2]{Anderson1973}, when $s$ moves along a small \mbox{semi-circular} indentation around $j\omega_0$ with radius $\epsilon$, i.e., $s\!=\!j\omega_0\!+\!\epsilon e^{j\theta},\theta\!\in\![-\frac{\pi}{2},\frac{\pi}{2}]$, we have
$x^*G(s)x= x^*K_0x\epsilon^{-k}e^{-jk\theta}+O(\epsilon^{-k+1})$,
where $K_0=\lim_{s\rightarrow j\omega_0}(s-j\omega_0)^k G(s)$ and $O(\epsilon^{-k+1})$ means a number in an order bounded by $\epsilon^{-k+1}$, for $x$ such that $\|x\|_2\!=\!1$ and $x^*K_0x\neq 0$. One can see that under the condition in the theorem statement, $k$ has to be $1$, otherwise two inequalities in (\ref{eq: gspt1}) will not hold simultaneously. This means that $\omega_0$ is at most a simple pole of any element of $G$. Then, on the semi-circular indentation,
\begin{align}
    G(s)\simeq \frac{K_0}{s-j\omega_0} \text{ and }H(s)\simeq H(j\omega_0)\label{indentapp}.
\end{align}
Since the inequality (\ref{eigine}) holds at both end points of the semi-circular indentation, in view of (\ref{indentapp}), we have
\begin{align*}
-\pi<\angle \lambda_i\left(\frac{K_0H(j\omega_0)}{s-j\omega_0}\right)<\pi
\end{align*}
for all nonzero $\lambda_i(\frac{K_0H(j\omega_0)}{s-j\omega_0})$ for all $s\!=\!j\omega_0+\epsilon e^{j\theta},\theta\!\in\![-\frac{\pi}{2},\frac{\pi}{2}]$ on the semi-circular indentation for a sufficiently small $\epsilon$.

The above analysis shows that $\det[I+G(s)H(s)]\neq 0$ for all points on the indented imaginary axis and the closed paths formed by the eigenloci of $GH$ along the indented imaginary axis do not encircle $-1$.
Next we will show that $j\omega_0$ is not a pole of $G\#H$. The closed-loop stability then follows from the generalized Nyquist criterion.

The system $G(s)$ can be written as
$\displaystyle G(s)=\frac{K_0}{s-j\omega_0}+G_0(s)$,
where $K_0$ is semi-sectorial and $G_0(s)$ is analytic at $j\omega_0$. Then, there exists a unitary matrix $U$ such that $U^*K_0U\!\!=\!\!\begin{bmatrix}
\tilde{K}_0& \\ & 0\end{bmatrix}$, where $\tilde{K}_0$ is nonsingular. Let $\tilde{G}(s)\!=\!U^*G(s)U$, $\hat{G}(s)\!=\!U^*G_0(s)U$, and $\tilde{H}(s)\!=\!U^*H(s)U$. Partitioning $\hat{G}(s)$ and $\tilde{H}(s)$ into
\[
\hat{G}(s)\!=\!\begin{bmatrix}
\hat{G}_{11}(s) & \hat{G}_{12}(s)\\ \hat{G}_{21}(s) & \hat{G}_{22}(s)\end{bmatrix} \text{ and } \tilde{H}(s)\!=\!\begin{bmatrix}
\tilde{H}_{11}(s) & \tilde{H}_{12}(s)\\ \tilde{H}_{21}(s) & \tilde{H}_{22}(s) \end{bmatrix}
\]
with compatible dimensions, we have
\begin{equation}\label{eq:pfe}
\tilde{G}(s)=\begin{bmatrix}
\frac{\tilde{K}_0}{s-j\omega_0}+\hat{G}_{11}(s) & \hat{G}_{12}(s)\\ \hat{G}_{21}(s) & \hat{G}_{22}(s)
\end{bmatrix}.
\end{equation}
The systems $(I\!+\!G(s)H(s))^{-1}$ and $(I\!+\!\tilde{G}(s)\tilde{H}(s))^{-1}$ have the same poles. We will analyze the unstable zeros of $I+\tilde{G}(s)\tilde{H}(s)$ by exploiting the coprime factorizations of $\tilde{G}(s), \tilde{H}(s)$ at $j\omega_0$. Similar to the usual coprime factorization theory \cite{Zhou}, a left-coprime fractorizaion of $\tilde{G}(s)$ at $j\omega_0$ is a factorization $\tilde{G}(s)=M(s)^{-1}N(s)$, where $M(s)$ and $N(s)$, analytic at $j\omega_0$, are left coprime in the sense that there exist $X(s)$ and $Y(s)$, analytic at $j\omega_0$, such that $M(s)X(s)+N(s)Y(s)=I$. Let
\begin{align*}
M(s)&=\begin{bmatrix}\frac{s-j\omega_0}{s+1}I &\\& I\end{bmatrix}, \
N(s)=\begin{bmatrix}
\frac{1}{s+1}\tilde{K}_0&\\&0\end{bmatrix} + \begin{bmatrix}
\frac{s-j\omega_0}{s+1}I &\\& I\end{bmatrix}\hat{G}(s), \\
Y(s)&=\begin{bmatrix} (1+j\omega_0)\tilde{K}_0^{-1} &\\&I
\end{bmatrix}, \
X(s)=-\hat{G}(s)Y(s)+I.
\end{align*}
Then it is easy to see that $M(s)X(s)\!+\!N(s)Y(s)\!=\!I$ and thus $\tilde{G}(s)\!=\!M(s)^{-1}N(s)$ is a left-coprime factorization at $j\omega_0$. Also, $\tilde{H}(s)=\tilde{H}(s)(I)^{-1}$ is obviously a right-coprime factorization of $\tilde{H}(s)$ at $j\omega_0$. By \cite[Lemma 5.10]{Zhou}, the unstable zeros of $I+\tilde{G}(s)\tilde{H}(s)$ are the unstable zeros of $M(s)I+N(s)\tilde{H}(s)$. Thus, it suffices to show that $M(j\omega_0)\!+\!N(j\omega_0)\tilde{H}(j\omega_0)$ is full rank. Let $A\!=\!\hat{G}(j\omega_0)$ and $B\!=\!\tilde{H}(j\omega_0)$. We have
\begin{align*}
M(j\omega_0)+N(j\omega_0)\tilde{H}(j\omega_0)
=\begin{bmatrix}
\frac{1}{j\omega_0+1}\tilde{K}_0B_{11} & \frac{1}{j\omega_0+1}\tilde{K}_0B_{12} \\ A_{21}B_{11}+A_{22}B_{21}& I+A_{21}B_{12}+A_{22}B_{22}
\end{bmatrix},
\end{align*}
which is of full rank if $\frac{\tilde{K_0}B_{11}}{j\omega_0+1}$ and its Schur complement, i.e., $I+A_{22}(B_{22}-B_{21}B_{11}^{-1}B_{12})$, are both of full rank. Since $B_{11}$ is a compression of sectorial $B$, it is nonsingular, which implies $\frac{\tilde{K_0}B_{11}}{j\omega_0+1}$ is full rank. Next we show $I+A_{22}(B_{22}-B_{21}B_{11}^{-1}B_{12})$ is full rank. It can be seen from (\ref{eq:pfe}) that $A_{22}$ is semi-sectorial. Also, $B_{22}-B_{21}B_{11}^{-1}B_{12}$ is the Schur complement of $B_{11}$ in $B$ and thus is sectorial \cite{WCKQ2020}.
In view of \eqref{eq: gspt1}, we have
\begin{align*}
\overline{\phi}(A_{22})+ \overline{\phi}(B_{22}-B_{21}B_{11}^{-1}B_{12}) <\pi, \ \ \
\underline{\phi}(A_{22})+ \underline{\phi}(B_{22}-B_{21}B_{11}^{-1}B_{12}) >-\pi,
\end{align*}
which implies that $I\!+\!A_{22}(B_{22}\!-\!B_{21}B_{11}^{-1}B_{12})$ is full rank. This completes the proof.
\end{proof}

This theorem subsumes a well-known version of passivity theorem \cite{LiuYao2016,Vidyasagar1981}, which states that $G\#H$ is stable if $G$ is positive real and $H$ is strongly positive real.


\subsection{Sectored real lemma for semi-stable semi-sectorial systems}
A state-space condition for a semi-stable semi-sectorial system $G$ can also be derived via the generalized KYP lemma. Let $j\Omega$ be the set of poles on the imaginary axis of $G$. Then, $\Phi_\infty(G) \subset [\alpha, \beta]$ with $\beta-\alpha \leq \pi$, if and only if $G(j\omega)$ has phases lie in $[-\frac{\pi}{2}+\eta_1,\frac{\pi}{2}+\eta_1]\cap[-\frac{\pi}{2}+\eta_2,\frac{\pi}{2}+\eta_2]$ for all $\omega\in[0,\infty]\backslash \Omega$, where $\eta_1=\alpha+\frac{\pi}{2}$ and $\eta_2=\beta-\frac{\pi}{2}$. This is equivalent to requiring that
\begin{align}
e^{-j\eta_i} G(j\omega)+e^{j\eta_i} G^*(j\omega)\geq 0, \;i=1,2,\label{ghsfdi}
\end{align}
hold for all $\omega\in[0,\infty]\backslash \Omega$. 

Applying the generalized KYP lemma, we can derive the following sectored real lemma for a semi-stable semi-sectorial $G$. The proof is similar to that of Theorem~\ref{thm: half-bound} and is omitted for brevity.

\begin{theorem}\label{thm:gsrl}
Let $G$ be a semi-stable semi-sectorial system with a minimal realization $\left[\begin{array}{c|c}A & B \\ \hline C & D\end{array} \right]$. Then $\Phi_\infty(G) \!\subset\! [\alpha, \beta]$, $ \beta - \alpha \leq \pi$, if and only if there exist Hermitian matrices $X_i, Y_i, i=1,2$, such that
\begin{align*}
Y_i\geq0, \begin{bmatrix}A&B\\I&0\end{bmatrix}'\!\begin{bmatrix}0&X_i+jY_i\\X_i-jY_i&0\end{bmatrix}\!\begin{bmatrix}A&B\\I&0\end{bmatrix}\!+\!M_i\leq0,
\end{align*}
where
\begin{align*}
M_i=\begin{bmatrix}0 & -e^{j\eta_i}C' \\ -e^{-j\eta_i}C & -e^{-j\eta_i}D-e^{j\eta_i}D' \end{bmatrix}, i=1,2.
\end{align*}
\end{theorem}

When $\alpha=-\frac{\pi}{2}$ and $\beta=\frac{\pi}{2}$, we have $\eta_1=\eta_2=0$. Then, the inequalities (\ref{ghsfdi}) reduce to
$G(j\omega)+G^*(j\omega)\geq 0$
for all $\omega\in[0,\infty]\backslash \Omega$. This means that $\Phi_{\infty}(G)\subset[-\frac{\pi}{2},\frac{\pi}{2}]$ if and only if $G$ is positive real. In this case, the two LMIs in Theorem \ref{thm:gsrl} reduce to one LMI
$\begin{bmatrix}A'X+XA&XB-C'\\B'X-C&-D-D'\end{bmatrix}\leq 0$,
which is the one in the positive real lemma \cite[Lemma 4]{Kottenstette}.

\section{Conclusion} \label{ending}
In this paper, we define the phase response of a MIMO LTI system. 
The newly defined phase concept enables the development of a phase theory of MIMO systems, complementing the magnitude-based system theory. A highlight of the paper is in the formulation of a small phase theorem, which is a counterpart of the well-known small gain theorem. We also derive a sectored real lemma for phase computation, which serves a counterpart of the bounded real lemma. In addition, we discuss the time-domain interpretation of the phase concept of MIMO systems.

This paper focuses on the analysis of MIMO systems based on the phase responses. A parallel study has been carried out for discrete-time LTI MIMO
systems \cite{MCQ2022}. A mixed gain/phase analysis by combining gain and phase information is under current investigation. The study of respective synthesis problems is also ongoing. The phase study has also been extended to nonlinear systems \cite{nonlinearphase}. 

\section*{Acknowledgment}
The authors would like to thank Jie Chen of City University of Hong Kong, Shinji Hara of Tokyo Institute of Technology, Chao Chen, Xin Mao, Qiangsheng Gao of Hong Kong University of Science and Technology, Axel Ringh of University of Gothenburg and Chalmers University of Technology, Di Zhao of Tongji University for helpful discussions.

\vspace{-10pt}

\bibliographystyle{siam}
\bibliography{ctphase,dynamic_v2}
\end{document}